\documentclass[journal]{IEEEtran}

\usepackage[cmex10]{amsmath}
\usepackage{amssymb,graphicx,enumerate,verbatim,xcolor}
\usepackage{xfrac}
\usepackage{tikz}
\usepackage{subcaption}
\usetikzlibrary{arrows,positioning,fit,backgrounds}
\usetikzlibrary{calc}
\usetikzlibrary{decorations.shapes}
\usetikzlibrary{shapes.geometric}

\newcommand{\eps}{\varepsilon}
\newcommand{\ovr}{\overline}

\newcommand{\xbar}{\overline{x}}
\newcommand{\Ebb}{\mathbb{E}}
\newcommand{\Nbb}{\mathbb{N}}
\newcommand{\Pbb}{\mathbb{P}}
\newcommand{\Rbb}{\mathbb{R}}
\newcommand{\Acal}{\mathcal{A}}
\newcommand{\Bcal}{\mathcal{B}}
\newcommand{\Ccal}{\mathcal{C}}

\newcommand{\Fcal}{\mathcal{F}}

\newcommand{\Kcal}{\mathcal{K}}
\newcommand{\Pcal}{\mathcal{P}}
\newcommand{\Rcal}{\mathcal{R}}
\newcommand{\Scal}{\mathcal{S}}
\newcommand{\Tcal}{\mathcal{T}}
\newcommand{\Ucal}{\mathcal{U}}
\newcommand{\Vcal}{\mathcal{V}}

\newcommand{\Xcal}{\mathcal{X}}
\newcommand{\Ycal}{\mathcal{Y}}
\newcommand{\Zcal}{\mathcal{Z}}

\newcommand{\avgcrit}{\mbox{\footnotesize \sc avg}}
\newcommand{\whpcrit}{\mbox{\footnotesize \sc whp}}
\newcommand{\mincrit}{\mbox{\footnotesize \sc min}}

\newtheorem{thm}{Theorem}
\newtheorem{cor}{Corollary}
\newtheorem{proposition}{Proposition}
\newtheorem{property}{Property}
\newtheorem{lemma}{Lemma}
\newtheorem{defn}{Definition}

\tikzstyle{arw}=[->,>=latex]
\tikzstyle{node}=[draw,rectangle,rounded corners]

\tikzstyle{col1}=[fill=red!80!]
\tikzstyle{col2}=[fill=green!80!black]
\tikzstyle{col3}=[fill=blue!80!]

\title{Rate-Distortion Theory for Secrecy Systems}
\author{Curt~Schieler and~Paul~Cuff%
\thanks{This work was supported in part by the National Science Foundation under Grants CCF-1116013 and CCF-1017431, and also by the Air Force Office of Scientific Research under Grant FA9550-12-1-0196. Portions of this paper were presented in \cite{Cuff2010globecom,Cuff2010allerton,Schieler2012,Cuff2013ita,Schieler2013}.}%
\thanks{The authors are with the Department of Electrical Engineering, Princeton University, Princeton, NJ, 08544 USA (email: schieler@princeton.edu; cuff@princeton.edu).}%
}

\begin{document}
\maketitle
\begin{abstract}
Secrecy in communication systems is measured herein by the distortion that an adversary incurs. The transmitter and receiver share secret key, which they use to encrypt communication and ensure distortion at an adversary. A model is considered in which an adversary not only intercepts the communication from the transmitter to the receiver, but also potentially has side information. Specifically, the adversary may have causal or noncausal access to a signal that is correlated with the source sequence or the receiver's reconstruction sequence. The main contribution is the characterization of the optimal tradeoff among communication rate, secret key rate, distortion at the adversary, and distortion at the legitimate receiver. It is demonstrated that causal side information at the adversary plays a pivotal role in this tradeoff. It is also shown that measures of secrecy based on normalized equivocation are a special case of the framework.
\end{abstract}


\section{Introduction}
\begin{figure}
\centering
 \begin{tikzpicture}
 [node distance=1cm,minimum width=1cm,minimum height =.75 cm]
  \node[rectangle,minimum width=5mm] (source) {Source $X$};
  \node[node] (alice) [right =7mm of source] {A};
  \node[node] (bob) [right =3cm of alice] {B};
  \node[coordinate] (dummy) at ($(alice.east)!0.5!(bob.west)$) {};
  \node[rectangle,minimum width=5mm] (xhat) [right =7mm of bob] {$\hat{X}$};
  \node[rectangle,minimum width=7mm] (key) [above =7mm of dummy] {secret key $K$};
  \node[node] (eve) [below =5mm of bob] {C};
  
  \draw [arw] (source) to (alice);
  \draw [arw] (alice) to node[minimum height=6mm,inner sep=0pt,midway,above]{message $M$} (bob);
  \draw [arw] (bob) to (xhat);
  \draw [arw] (key) to [out=180,in=90] (alice);
  \draw [arw] (key) to [out=0,in=90] (bob);
  \draw [arw,rounded corners] (dummy) |- (eve);
 \end{tikzpicture}
 \caption{\small The Shannon cipher system. Nodes A, B, and C are the transmitter, receiver, and eavesdropper, respectively.}
 \label{fig:scs}
 \end{figure}
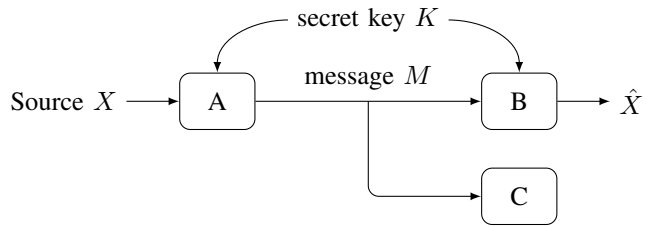

In ``Communication Theory of Secrecy Systems" \cite{Shannon1949}, Shannon regarded a communication system as perfectly secret if the source and the eavesdropped message are statistically independent. The secrecy system studied in \cite{Shannon1949} is referred to as the ``Shannon cipher system" and is depicted in Fig.~\ref{fig:scs}. A necessary and sufficient condition for perfect secrecy is that the number of secret key bits per source symbol exceeds the entropy of the source. When the amount of key is insufficient, one must relax the requirement of statistical independence and invite new measures of secrecy. 

One common way of measuring sub-perfect secrecy is with equivocation, the conditional entropy $H(X|M)$ of the source given the public message. The use of equivocation as a measure of secrecy was considered in the original work on the wiretap channel in \cite{Wyner1975wiretap} and \cite{Csiszar1978}, and it continues today. Although a distortion-based approach to secrecy might appear incomparable at first glance, it turns out that equivocation (when normalized by blocklength) becomes a special case of the framework developed here, under the proper choice of distortion measure.

In this work, we study an information-theoretic measure of secrecy that is directly inspired by rate-distortion theory. Whereas the objective in classical rate-distortion theory is to minimize a receiver's distortion for a given rate of communication, our goal is to maximize an eavesdropper's distortion for a given rate of secret key. If we relax the requirement of lossless communication in Shannon's cipher system, then our goal is to maximize an eavesdropper's distortion for a given secret key rate, communication rate, and distortion tolerance at the receiver. Although there are a variety of secrecy systems other than Shannon's cipher system (such as a wiretap channel \cite{Wyner1975wiretap} or distributed correlated sources 
 \cite{Maurer1993,Ahlswede1993}), this paper is concerned exclusively with settings involving shared secret key, a single discrete memoryless source, and a noiseless channel. Moreover, we focus on block codes in the regime of blocklength tending to infinity.

When distortion is used as a measure of secrecy, we are implicitly viewing an eavesdropper in the same way that one views a receiver in a standard rate-distortion setting -- as an active participant whose goal is to produce a sequence that is statistically correlated with the source sequence. Because he plays an active role, the eavesdropper is thought of as an adversarial entity. To ensure robustness, we will design the communication and encryption schemes against the worst-case adversarial strategy; that is, we wish to maximize the minimum distortion attainable by an adversary. 
 
 The study of information-theoretic secrecy via rate-distortion theory was initiated by Yamamoto in \cite{Yamamoto1988}, in which the rate-distortion region was characterized for the special setting in which no secret key is available. Later, in ``Rate-Distortion Theory for the Shannon Cipher System" \cite{Yamamoto1997}, Yamamoto considered the exact problem we have heretofore described, but only obtained an inner and outer bound on the achievable rate-key-distortion region.\footnote{The inner bound provided in \cite{Yamamoto1997} is precisely the region expressed in \eqref{lossyXregion} of this work.  Corollary~\ref{fourcases} shows that this performance is achievable even if additional information is available to the eavesdropper, but it is suboptimal for the problem at hand.  The outer bound in \cite{Yamamoto1997} makes use of two auxiliary variables, but with the appropriate selection can be shown to be equivalent to the trivial bound in \eqref{lossyNOregion}, which in fact we show to be achievable.  To show that the outer bound in \cite{Yamamoto1997} is trivial, the variables $U$ and $V$ can be selected as follows. Let $U$ be independent of $X$ and $Y$ and uniformly distributed on $\{1,\ldots,|\Xcal|\}$.  Let $V=U + X$ modulo $|\Xcal|$.} In this paper, we characterize the region; however, it is not our main focus. The following example serves to illustrate the care that should be exercised in a distortion-based approach to secrecy and motivates our primary investigation, which is centered around a salient feature of our model referred to as \emph{causal disclosure}.
 




\subsection{One-bit secrecy and causal disclosure}
Consider an $n$-bit i.i.d. source sequence $X^n~\triangleq~(X_1,\ldots,X^n)$ with $X_i \sim \text{Bern}(1/2)$. Suppose common randomness $K\sim \text{Bern}(1/2)$ is available to the transmitter and receiver; that is, there is one bit of shared secret key. Now suppose the transmitter uses $K$ to encrypt $X^n$ by transmitting the $n$-bit message $\widetilde{X}^n$, where $\widetilde{X}_i=X_i\oplus K$. In other words, he flips all of the bits of $X^n$ if $K=1$, otherwise he simply sends $X^n$. Upon intercepting the public message $\widetilde{X}^n$, the adversary produces a reconstruction $Z^n$ and incurs expected distortion $\Ebb\,\frac1n \sum_{i=1}^n d(X_i,Z_i)$, where $d(x,z)$ is a per-letter distortion measure. If $d(x,z)=\mathbf{1}\{x\!\neq\! z\}$, then an optimal strategy for the adversary is to simply set $Z^n=\widetilde{X}^n$, yielding an expected distortion of $1/2$. Observe that $1/2$ is also the maximum possible expected distortion that we could ever force on the adversary, regardless of the amount of secret key available! It appears as though we have maximized secrecy by only using one bit of secret key for an arbitrarily long $n$-bit source. However, this view is severely misleading because the adversary actually knows a great deal about $X^n$, namely that it is one of only two candidate sequences.  


This example demonstrates the potential fragility of using distortion to measure secrecy without recognizing the ramifications. For, although maximum secrecy (in the distortion sense) is attained, it vanishes altogether if the adversary views just one true bit of the source sequence (the bit allows him to determine whether or not to flip the $\widetilde{X}^n$ sequence). In general, the consequences of this example apply to the setting that Yamamoto considered in \cite{Yamamoto1997}. An arbitrarily small rate of secret key is enough to guarantee maximum distortion, but such secrecy is weak in the sense that even a small amount of additional knowledge (for example, observation of a few source symbols) is enough for the adversary to completely identify the source sequence.

The way that we strengthen a distortion-based approach to secrecy is through an assumption of causal disclosure, in which we design codes under the supposition that the adversary has noisy (or noiseless) access to the past behavior of the system. For example, in the one-bit secrecy example we might assume that the adversary produces the $i$th reconstruction symbol $Z_i$ based not only on the public message $M$, but also on the past source symbols $X^{i-1}$. Incidentally, such a modification to the standard rate-distortion theory setting does not change the theory, though it has a dramatic effect in this secrecy setting. Regardless of whether or not an adversary actually has access to such information, designing our encryption under the assumption that he does leads to a much more robust notion of secrecy. In particular, it is resistant to disruptions in secrecy like those exhibited in the example. Despite the ``pessimistic" nature of the causal disclosure assumption, we find that the optimal tradeoff between secret key and distortion in this regime is reasonable and not degenerate.

The assumption of causal disclosure is relevant not only for the sake of robustness, but also for its natural interpretations. In \cite{Cuff2010}, an alternative view of rate-distortion theory was introduced in which source and reconstruction sequences are regarded as sequences of actions in a distributed system. Communication is used to coordinate the receiver's actions with the transmitter's actions (which are given by nature). In this context, an adversary can be viewed as an active participant in the system who produces a sequence of actions. With this interpretation, it is not unrealistic to assume that the adversary could have causal access to the system behavior. Depending on where the adversary is intercepting communication, he might be able to view the past actions of the transmitter or receiver (or both) and produce his current action accordingly.

We find that optimal communication in this setting is not only fundamentally different than that of other source coding problems (often requiring a stochastic decoder), but in fact lends itself to a simple interpretation of injecting artificial memoryless noise into the adversary's received signal.

\subsection{Organization}
The content of this paper is as follows. In Section~\ref{sec:prelim}, we describe the problem setup. In Section~\ref{sec:onebit}, we present a generalized version of the one-bit secrecy example in which there is no assumption of causal disclosure. In Section~\ref{sec:mainresult}, we state our main result, Theorem~\ref{mainthm}, in which causal disclosure is a primary assumption. Theorem~\ref{mainthm} describes the optimal relationship among the communication rate, secret key rate, and distortion at the legitimate receiver and adversary. Section~\ref{sec:mainresult} also establishes a number of relevant corollaries to Theorem~\ref{mainthm} and provides several concrete examples of the corresponding information-theoretic tradeoff regions. In Section~\ref{sec:equivocation}, we demonstrate how normalized equivocation arises as a special case of the causal disclosure framework. In Section~\ref{sec:achievability}, we give the achievability proof of Theorem~\ref{mainthm}. The proof uses a stochastic ``likelihood encoder" that enables tractable analysis when combined with a ``soft covering lemma". Afterward, we discuss several important properties and implications of the optimal communication scheme used in the proof. Section~\ref{sec:converse} provides the converse proof of Theorem~\ref{mainthm}. In Section~\ref{sec:relatedresults}, we consider some settings with noncausal disclosure that are not subsumed by Theorem~\ref{mainthm}, but that can be proved similarly. Lastly, Section~\ref{sec:delay} gives results for settings involving causal disclosure with delay greater than one. 

\section{Preliminaries}
\label{sec:prelim}
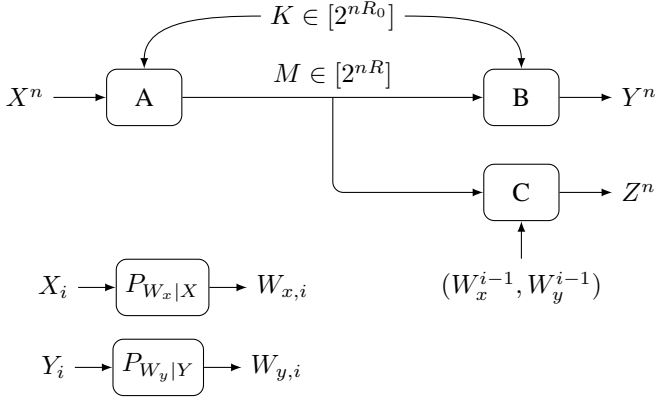
\begin{figure}
\centering
 \begin{tikzpicture}
 [node distance=1cm,minimum width=1cm,minimum height =.75 cm]
  \node[rectangle,minimum width=5mm] (source) {$X^n$};
  \node[node] (alice) [right =7mm of source] {A};
  \node[node] (bob) [right =4cm of alice] {B};
  \node[coordinate] (dummy) at ($(alice.east)!0.5!(bob.west)$) {};
  \node[rectangle,minimum width=5mm] (xhat) [right =7mm of bob] {$Y^n$};
  \node[rectangle,minimum width=7mm] (key) [above =7mm of dummy] {$K\in[2^{nR_0}]$};
  \node[node] (eve) [below =5mm of bob] {C};
  \node[rectangle,minimum width=5mm] (zn) [right =7mm of eve] {$Z^n$};
  \node[rectangle] (side) [below=5mm of eve] {$(W_x^{i-1},W_y^{i-1})$};
  
  \node[rectangle,minimum width=5mm] (xinput) at (source.south east |- side.west) {$X_i$};
  \node[node] (chx) [right =5mm of xinput] {$P_{W_x|X}$};
  \node[rectangle,minimum width=5mm] (xoutput) [right =5mm of chx] {$W_{x,i}$};
  
  \node[rectangle,minimum width=5mm] (yinput) [below =3mm of xinput] {$Y_i$};
  \node[node] (chy) [right =5mm of yinput] {$P_{W_y|Y}$};
  \node[rectangle,minimum width=5mm] (youtput) [right =5mm of chy] {$W_{y,i}$};

  \draw [arw] (source) to (alice);
  \draw [arw] (alice) to node[minimum height=6mm,inner sep=0pt,midway,above]{$M\in[2^{nR}]$} (bob);
  \draw [arw] (bob) to (xhat);
  \draw [arw] (key) to [out=180,in=90] (alice);
  \draw [arw] (key) to [out=0,in=90] (bob);
  \draw [arw,rounded corners] (dummy) |- (eve);
  \draw [arw] (eve) to (zn);
  \draw [arw] (side) to (eve);
  \draw [arw] (xinput) to (chx);
  \draw [arw] (chx) to (xoutput);
  \draw [arw] (yinput) to (chy);
  \draw [arw] (chy) to (youtput);
 \end{tikzpicture}
 \caption{\small Nodes A and B use secret key $K$ and public communication $M$ to coordinate against an adversarial Node C. At each step $i$, Node C can view the past behavior of the system, $(W_x^{i-1},W_y^{i-1})$, where $W_x^n$ is the output of a memoryless channel $\prod P_{W_x|X}$ with input $X^n$, and $W_y^n$ is the output of a memoryless channel $\prod P_{W_y|Y}$ with input $Y^n$.}
 \label{fig:model}
 \end{figure}
The communication system model used throughout is shown in Fig.~\ref{fig:model}. The transmitting node, Node A, observes an i.i.d.\ source sequence $X^n\triangleq(X_1,\ldots,X_n)$, where $X_i$ is distributed according to $P_X$. Nodes A and B share a source of common randomness $K\in\{1,\ldots,2^{nR_0}\}$, referred to as secret key, that is uniformly distributed and independent of $X^n$. Based on the source block $X^n$ and the secret key $K$, Node A transmits a message $M\in\{1,\ldots,2^{nR}\}$ that is received without loss by Nodes B and C. Once $M$ is delivered, all three nodes sequentially produce actions: in the $i$th step, Nodes A, B and C produce $X_i$, $Y_i$, and $Z_i$, respectively. Note that Node A has no control over his actions; they are simply given by $X^n$. At the other end, Node B produces $Y_i$ based on the pair $(M,K)$ and the adversarial Node C produces $Z_i$ based on $M$ and his observation of the past behavior of the system, $(W_x^{i-1},W_y^{i-1})$. At each step, the joint actions of the players incur a value $\pi(x,y,z)$, which represents symbol-wise payoff; the block-average payoff is given by 
\begin{equation}
\frac1n \sum_{i=1}^n \pi(X_i,Y_i,Z_i).
\end{equation}
 Nodes A and B want to cooperatively maximize payoff, while Node C wants to minimize payoff through his actions $Z^n$. This payoff function can take the role of distortion incurred by Node C, corresponding to the secrecy metric described in the introduction. Note that instead of evaluating secrecy and coordination separately, which could be done with two payoff functions $\pi_1(x,y)$ and $\pi_2(x,y,z)$, we have unified them in a single function $\pi(x,y,z)$. Of course, the use of multiple payoff functions does have its own merits, and the results extend readily.

In Fig.~\ref{fig:model}, we depict noisy causal disclosure by $(W_x^{i-1},W_y^{i-1})$, where $W_x^n$ is the output of a memoryless channel $\prod_{i=1}^n P_{W_x|X}$ with input $X^n$, and $W_y^n$ is the output of a memoryless channel $\prod_{i=1}^n P_{W_y|Y}$ with input $Y^n$. Modeling the side information in this way covers a variety of scenarios. For example, if $P_{W_x|X}$ and $P_{W_y|Y}$ are identity channels, resulting in $(W_x,W_y)=(X,Y)$, then the adversary has full causal access $(X^{i-1},Y^{i-1})$. This is the strongest definition of secrecy in the causal disclosure framework and leads to the design of a thoroughly robust secrecy system. If $(W_x,W_y)=(\emptyset,\emptyset)$, then the adversary is completely blind to the past and only views the public message $M$; this is the setting of \cite{Yamamoto1997}, which does not include causal disclosure. 

We remark that other strong security definitions involving side information leaks to the adversary can be found in \cite{Maurer2012}, for example.



Throughout, we assume that the alphabets $\Xcal$, $\Ycal$, and $\Zcal$ are finite. We denote the set $\{1,\ldots,m\}$ by $[m]$ and use $\Delta_{\Acal}$ to denote the probability simplex of distributions with alphabet $\Acal$. The notation $X \perp Y$ indicates that the random variables $X$ and $Y$ are independent, and $X-Y-Z$ indicates a markov chain relationship.
\begin{defn}
An $(n,R,R_0)$ code consists of an encoder $f~:~\Xcal^n\times [2^{nR_0}]\rightarrow [2^{nR}]$ and a decoder $g: [2^{nR}]\times [2^{nR_0}]\rightarrow \Ycal^n$. More generally, we allow a stochastic encoder $P_{M|X^n,K}$ and a stochastic decoder $P_{Y^n|M,K}$. An $(n,R,R_0)$ code is said to have blocklength $n$, communication rate $R$, and secret key rate $R_0$.
\end{defn}

Permitting stochastic decoders that use local randomization is crucial (in contrast to Wyner's wiretap channel, in which a stochastic \emph{encoder} is needed). On the other hand, it is likely that the optimal encoder can be a deterministic function of the message and key, but this has not been shown. The proof of our main result uses a stochastic encoder and stochastic decoder.


Nodes A and B use an $(n,R,R_0)$ code to coordinate against Node C. To ensure robustness, we consider the payoff that can be assured against the worst-case adversary, i.e., the max-min payoff. There are several ways to define the payoff criterion for a block, and we consider three: expected payoff, assured payoff, and symbol-wise minimum payoff. To distinguish among the three criteria, we use the monikers {\sc avg}, {\sc whp}, and {\sc min}, respectively.

\begin{defn}
\label{defnachievability}
 Fix a source distribution $P_X$, a symbol-wise payoff function $\pi:\Xcal\times\Ycal\times\Zcal\rightarrow \Rbb$, and causal disclosure channels $P_{W_x|X}$ and $P_{W_y|Y}$. For simplicity, denote the pair $(W_x^n,W_y^n)$ by $W^n$. The triple $(R,R_0,\Pi)$ is achievable if there exists a sequence of $(n,R,R_0)$ codes such that
\begin{itemize}
  \item Under the {\sc avg} criterion (expected payoff):
  \begin{equation}
    \liminf_{n\rightarrow\infty}\min_{\{P_{Z_i|M,W^{i-1}}\}_{i=1}^n}\Ebb\,\frac1n \sum_{i=1}^n \pi(X_i,Y_i,Z_i)\geq\Pi.
  \end{equation}
  \item Under the {\sc whp} criterion (assured payoff): 
  \begin{equation}
  \lim_{n\rightarrow\infty} \min_{\{P_{Z_i|M,W^{i-1}}\}_{i=1}^n} \Pbb \Big[\frac1n \sum_{i=1}^n \pi(X_i,Y_i,Z_i)\geq\Pi \Big] = 1.
  \end{equation}
  \item Under the {\sc min} criterion (symbol-wise minimum payoff):
  \begin{equation}
  \liminf_{n\rightarrow\infty}\min_{i\in[n]} \min_{P_{Z|M,W^{i-1}}} \Ebb\,\pi(X_i,Y_i,Z) \geq \Pi.
  \end{equation}
 \end{itemize}
 Under the {\sc whp} criterion, the range of $\pi(x,y,z)$ is extended to include $-\infty$ so that lossless communication settings can be recovered.
\end{defn}
Several remarks concerning the preceding definitions are in order.
\begin{enumerate}
\item Although {\sc whp} and {\sc min} are incomparable, they are both stronger than {\sc avg}. However, it will be shown that all three criteria give rise to the same optimal tradeoff region.
\item In each of the criteria, we allow the adversary to employ his best set of probabilistic strategies $\{P_{Z_i|M,W^{i-1}}\}_{i=1}^n$ that minimize payoff. However, since expectation is linear in $P_{Z_i|M,W^{i-1}}$ for all $i$, the expectation is minimized by extreme points of the probability simplex; thus, we can assume that Node C uses a set of deterministic strategies, $\{z_i(m,w^{i-1})\}_{i=1}^n$.
\item It is assumed (although not explicit in the notation) that the adversary has full knowledge of the source distribution and the code that Nodes A and B use.
\item The optimal payoff does not increase if Node B is given direct causal access to Nodes A and C (i.e., if the decoder is given by $\{P_{Y_i|M,K,X^{i-1},Z^{i-1}}\}_{i=1}^n$ instead of simply $P_{Y^n|M,K}$). This is shown in Section~\ref{sec:converse} in the converse proof of the main result.
\end{enumerate}

\begin{defn}
\label{defnregion}
The rate-payoff region $\Rcal_{\avgcrit}$ is the closure of achievable triples $(R,R_0,\Pi)$ under payoff criterion {\sc avg}. Regions $\Rcal_{\whpcrit}$ and $\Rcal_{\mincrit}$ are defined in the same way.
\end{defn}

\section{One-bit secrecy, generalized}
\label{sec:onebit}
In this section, we expand on the scenario in which lossless communication is required between Nodes A and B and there is no causal disclosure of the system behavior to Node C. This is Yamamoto's setting in \cite{Yamamoto1997}. Although the result of this section is a special case of the main result in Theorem~\ref{mainthm}, it is an illustrative starting point. 

For lossless communication, an additional achievability criterion is required, as stated below. Since $X^n$ must equal $Y^n$ with high probability, the payoff function is of the form $\pi(x,z)$. Thus, the achievability criteria for $(R,R_0,\Pi)$ under the {\sc min} payoff criterion (which is stronger than the {\sc avg} criterion) are that 
\begin{equation}
\label{proberror}
\lim_{n\rightarrow\infty}\Pbb[X^n\neq Y^n]=0
\end{equation}
and
\begin{equation}
\label{propcriterion}
\liminf_{n\rightarrow\infty}\min_{i\in[n]} \min_{z(m)} \Ebb\,\pi(X_i,z(M)) \geq \Pi.
\end{equation}

\begin{proposition}
\label{secrecyischeap}
Fix $P_X$ and $\pi(x,z)$. If lossless communication is required and there is no causal disclosure, then $\Rcal_{\mincrit}$, the rate-payoff region under payoff criteria {\sc avg} and {\sc min} is equal to
 \begin{equation}
 \label{propositionregion}
 \left\{
 \begin{IEEEeqnarraybox}[][c]{rCl}
 \IEEEeqnarraymulticol{3}{l}{
 (R,R_0,\Pi): \vspace{2pt}
 }\\
 R &\geq& H(X)\\
 R_0 &\geq& 0\\
 \Pi &\leq& \min_z \Ebb\, \pi(X,z)
 \end{IEEEeqnarraybox}
\right\}.
\end{equation}
\end{proposition}

Thus, any positive rate of secret key\footnote{Note that $R_0=0$ is only included in Proposition~\ref{secrecyischeap} because we defined the region as the \emph{closure} of achievable triples. Furthermore, we remark that $R_0=0$ refers to a vanishing rate of secret key and is not the same as the absence of key.} guarantees maximum secrecy (in the distortion sense), as Node C can achieve $\min_z \Ebb\, \pi(X,z)$ by only knowing the source statistics. In fact, we now prove that each point in \eqref{propositionregion} can be achieved with key size $\Kcal=[n]$ instead of $\Kcal=[2^{nR_0}]$. This shows that even if the number of secret key bits is sublinear in the blocklength (in this case, $\log n$), one can still force the eavesdropper to incur the maximum distortion.\footnote{In \cite{Schieler2012}, we show that an arbitrarily slow rate of increase is sufficient, even slower than $\log n$, under the {\sc avg} criterion.} As in the example of one-bit secrecy, such guarantees are shattered if even a small amount of source information is available to the adversary.

The following lemma is useful for the payoff analysis.

\begin{lemma}
\label{suffstat}
Let $P_{XYZ}$ be a markov chain $X-Y-Z$, and $f$ an arbitrary function. Then
\begin{equation}
\min_{g(x,y)}\Ebb\,f(g(X,Y),Z)=\min_{g(y)}\Ebb\,f(g(Y),Z).
\end{equation}
\end{lemma}

\begin{IEEEproof}
We have
\begin{IEEEeqnarray}{rCl}
\IEEEeqnarraymulticol{3}{l}{
\nonumber\min_{g(x,y)}\Ebb\,f(g(X,Y),Z)
}\\
\vspace{-10pt} \nonumber &=& \min_{g(x,y)}\sum_{x,y}P_{X,Y}(x,y)\Ebb[f(g(X,Y),Z)|(X,Y)=(x,y)] \\
&& \vspace{5pt}\\
&=& \sum_{x,y}P_{X,Y}(x,y) \min_{g} \Ebb[f(g,Z)|(X,Y)=(x,y)]\\
&\stackrel{(a)}{=}& \sum_{x,y}P_{X,Y}(x,y) \min_{g} \Ebb[f(g,Z)|Y=y]\\
&=& \min_{g(y)}\Ebb\,f(g(Y),Z),
\end{IEEEeqnarray}
where (a) follows from the markovity assumption.
\end{IEEEproof}

Now we prove Proposition~\ref{secrecyischeap}.
\begin{IEEEproof}[Proof of Proposition~\ref{secrecyischeap}]
\emph{Converse.} By the converse to the lossless source coding theorem, if \eqref{proberror} holds then we must have $R\geq H(X)$. To see that the payoff never exceeds $\min_z \Ebb\,\pi(X,z)$, observe that the adversary can always let $Z^n$ equal $(z^*,\ldots,z^*)$, where
\begin{equation}
z^*=\arg\!\min_z \Ebb\,\pi(X,z). 
\end{equation}
Note that this converse argument holds for all three payoff criteria.

\emph{Achievability.} 
Let $\eps>0$. Denote the empirical distribution (also referred to as the type) of a sequence $x^n$ by $P_{x^n}$:
\begin{equation}
P_{x^n}(x)=\frac1n \sum_{i=1}^n \mathbf{1}\{x_i=x\}.
\end{equation}
The set of $\eps$-typical sequences is defined as
\begin{equation}
\label{typicalsetdefn}
\Tcal_\eps^n\triangleq\{x^n:|P_{x^n}(x)-P_X(x)|<\eps P_X(x),\forall x\in\Xcal\}.
\end{equation}

To communicate, Nodes A and B use the set of $\eps$-typical sequences as their codebook, just as in the standard proof of the lossless source coding theorem. If the source sequence $X^n$ is typical, then the index of that codeword is the (pre-encrypted) message; if the source sequence is not typical, an arbitrary index is selected. Due to familiar properties of the size and probability of the typical set, the rate of communication is $(1+\eps)H(X)$ and the probability of error is
\begin{equation}
\label{proberror2}
\Pbb[X^n \neq Y^n] < \eps
\end{equation}
for large enough $n$. 

The message will be encrypted using common randomness $K\sim\text{Unif}[n]$; this implies that the rate of secret key approaches zero as blocklength tends to infinity.  In order to encrypt, we first partition $\Tcal_\eps^n$ into bins of size $n$ (in a manner specified shortly), and use $K$ to apply a one-time pad to the location of the source sequence $X^n$ within the appropriate bin. More precisely, the encoder operates as follows: if $X^n$ is $\eps$-typical and is the $L$th sequence in the $J$th bin, then transmit the message $M=(J,L\oplus K)$, where $\oplus$ indicates addition modulo $n$. By encrypting in this manner, the adversary knows which bin $X^n$ lies in (bin $J$), but does not know which of those $n$ sequences it is, because $L$ is independent of $L\oplus K$. Using the secret key, Node B can recover both $J$ and $L$ and produce the corresponding sequence.

The partitioning of $\Tcal_{\eps}^n$ is done according to the following equivalence relation:
\begin{equation}
\label{equivrelation}
x^n\sim y^n \mbox{ if } x^n \mbox{ is a cyclic permutation of } y^n.
\end{equation}
Although the resulting partition can contain bins of size less than $n$, the number of such bins is small enough that we can ignore them without affecting the communication rate or \eqref{proberror2}. Thus, we assume that partitioning $\Tcal_{\eps}^n$ yields only bins of size $n$.  Due to \eqref{equivrelation}, it can be readily shown that each bin of size $n$ has the following property.
\begin{property}
\label{binproperty}
View the $j$th bin (denoted by $b_j$) as an $n\times n$ matrix whose columns are formed from the sequences in the bin. Then every row and column of the matrix has the same empirical distribution (denoted by $P_j$) and hence every row has the same probability (denoted by $\alpha_j$) under the source distribution $\prod_{i=1}^n P_X(x_i)$.
\end{property}
 
This property is the crux of the proof; we offer the following intuition for why it implies that the eavesdropper suffers maximal distortion. The eavesdropper knows which bin $X^n$ lies in, but does not know where it lies in the bin. Because of how we partitioned $\Tcal_{\eps}^n$, the eavesdropper's uncertainty is spread uniformly over the bin. To estimate $X_i$, the eavesdropper consults the $i$th row of the bin; however, Property~\ref{binproperty} ensures that the empirical distribution of this row matches the type of the sequences in the bin, which in turn approximates the source distribution $P_X$ (due to typicality). Therefore, the eavesdropper's estimate of $X_i$ is based on no more than the original source statistics, which means that he suffers maximal distortion.

We now analyze the distortion precisely. For sufficiently large $n$, we have for all $i\in[n]$ that
\begin{IEEEeqnarray}{rCl}
\IEEEeqnarraymulticol{3}{l}{
\nonumber\min_{z(m)} \Ebb\, d(X_i,z(M))
}\\
\quad &=& \min_{z(j,l)} \Ebb\, d(X_i,z(J,L\oplus K))\\
&\stackrel{(a)}{=}& \min_{z(j)} \Ebb\, d(X_i,z(J))\\
&=& \min_{z(j)} \sum_j \sum_{x^n\in b_j} p(x^n) d(x_i,z(j))\\
&\stackrel{(b)}{=}& \sum_j \alpha_j \min_z \sum_{x^n\in b_j} d(x_i,z)\\
&\stackrel{(c)}{=}& \sum_j \alpha_j \min_z \sum_{x\in\Xcal} n P_j(x)d(x,z)\\
&\stackrel{(d)}{\geq}& \sum_j n\alpha_j \min_z \sum_{x\in\Xcal} (1-\eps) P_X(x)d(x,z)\\
&=&\Pbb[X^n\in \Tcal_{\eps}^n](1-\eps)\min_z \Ebb\,d(X,z)\\
&\geq&(1-\eps)^2\min_z \Ebb\,d(X,z),
\end{IEEEeqnarray}
where (a) is due to $(X_i,J) \perp (L\oplus K)$ and Lemma~\ref{suffstat}, (b) and (c) are due to Property~\ref{binproperty}, and (d) follows from the definition of $\Tcal_\eps^n$. Thus, we have \eqref{propcriterion}.
\end{IEEEproof}
\subsection*{Discussion}
Suppose Nodes A and B use the binning scheme just described in the proof of Proposition~\ref{secrecyischeap} to achieve maximum secrecy. What if, instead of eavesdropping only the public message, the adversary is also able to view the past behavior of the system, namely $X^{i-1}$? Because of the structure of each bin (i.e., Property \ref{binproperty}), knowledge of just the first symbol, $X_1=x_1$, is enough for the adversary to narrow down the size of the list of candidate source sequences from $n$ to approximately $nP_{X}(x_1)$. One can see that the adversary will be able to determine the true sequence quickly, well before the end of the block. In this manner, the adversary can take advantage of the causal disclosure to force the payoff to take on its minimum value instead of its maximum value. In general, causal disclosure benefits an adversary and gives rise to a nontrivial tradeoff between secret key and payoff. We remark that one of the key elements in the proof of the main result is that the benefits of causal disclosure can be voided if the right amount of secret key is available. In fact, it will become evident in Section~\ref{sec:achievability} that using secret key to sterilize the causal disclosure gives rise to the optimal tradeoff of secret key and payoff.

\section{Main Result}
\label{sec:mainresult}

Our main result is the following.
\begin{thm}
\label{mainthm}
Fix $P_X$, $\pi(x,y,z)$, and causal disclosure channels $P_{W_x|X}$ and $P_{W_y|Y}$. Then $\Rcal_{\avgcrit}$, the closure of achievable $(R,R_0,\Pi)$ under payoff criterion {\sc avg}, is equal to

 \begin{equation}
 \label{mainthmregion}
 \bigcup_{W_x-X-(U,V)-Y-W_y}\left\{
 \begin{IEEEeqnarraybox}[][c]{rCl}
 \IEEEeqnarraymulticol{3}{l}{
 (R,R_0,\Pi): \vspace{2pt}
 }\\
 R &\geq& I(X;U,V)\\
 R_0 &\geq& I(W_xW_y;V|U)\\
 \Pi &\leq& \min_{z(u)} \Ebb\, \pi(X,Y,z(U))
 \end{IEEEeqnarraybox}
\right\},
\end{equation}

where $|\Ucal|\leq |\Xcal|+2$ and $|\Vcal|\leq |\Xcal||\Ycal|(|\Xcal|+2)+1$. Furthermore, 
\begin{equation}
\label{equivregion}
\Rcal_{\avgcrit}=\Rcal_{\whpcrit}=\Rcal_{\mincrit}.
\end{equation}

\end{thm}
We remark that the convexity of $\Rcal_{\avgcrit}$ and $\Rcal_{\whpcrit}$ can be shown from Definitions~\ref{defnachievability} and~\ref{defnregion} by using a standard time-sharing argument. By \eqref{equivregion}, $\Rcal_{\mincrit}$ is also a convex set.

We now elaborate on several corollaries to Theorem~\ref{mainthm} that are obtained through different choices of the causal disclosure channels $P_{W_x|X}$ and $P_{W_y|Y}$. To begin, we consider scenarios in which lossless communication is required between Nodes A and B.
\subsection{Lossless communication}
In the following, we require $X^n$ to equal $Y^n$ with high probability. That is, we introduce into Definition~\ref{defnachievability} the additional constraint
\begin{equation}
\label{losslessconstraint}
\lim_{n\rightarrow\infty}\Pbb[X^n\neq Y^n]=0.
\end{equation}
Conveniently, \eqref{losslessconstraint} can be ensured by considering payoff criterion {\sc whp} with a payoff function $\pi(x,y,z)$ that evaluates to $-\infty$ when $x\neq y$.
\begin{cor}
\label{losslessW}
Fix $P_X$, $\pi(x,z)$, and causal disclosure channel $P_{W_x|X}$. If lossless communication is required (i.e., \eqref{losslessconstraint} is imposed), then the rate-payoff region $\Rcal_{\whpcrit}$ is equal to
 \begin{equation}
 \label{losslessWregion}
 \bigcup_{U-X-W_x}\left\{
 \begin{IEEEeqnarraybox}[][c]{rCl}
 \IEEEeqnarraymulticol{3}{l}{
 (R,R_0,\Pi): \vspace{2pt}
 }\\
 R &\geq& H(X)\\
 R_0 &\geq& I(W_x;X|U)\\
 \Pi &\leq& \min_{z(u)} \Ebb\, \pi(X, z(U))
 \end{IEEEeqnarraybox}
\right\}.
\end{equation}
\end{cor}
\begin{IEEEproof}
Define a payoff function
\begin{equation}
\ovr{\pi}(x,y,z)\triangleq\begin{cases}
\pi(x,z) & \mbox{if } x=y\\
-\infty & \mbox{if } x\neq y.
\end{cases}
\end{equation}
When $\Pi>-\infty$, it is easily verified that
\begin{equation}
\lim_{n\to\infty}\Pbb\Big[\frac1n \sum_{i=1}^n \ovr{\pi}(X_i,Y_i,Z_i) \geq \Pi -\eps\Big] = 1
\end{equation}
if and only if both of the following hold:
\begin{IEEEeqnarray}{l}
\lim_{n\to\infty}\Pbb[X^n = Y^n] = 1\\
\lim_{n\to\infty}\Pbb\Big[\frac1n \sum_{i=1}^n \pi(X_i,Z_i) \geq \Pi -\eps\Big] = 1.
\end{IEEEeqnarray}
Thus, $\Rcal_{\whpcrit}$ (the region we seek the characterize) is obtained by invoking Theorem~\ref{mainthm} with $W_y=\emptyset$. However, we want to simplify the region further. Denoting the region in \eqref{losslessWregion} by $\Scal$, we now show that $\Rcal_{\whpcrit}=\Scal$.  

Note that when $\Pi>-\infty$, we have
\begin{equation}
-\infty < \Pi \leq \min_{z(u)} \Ebb\,\ovr{\pi}(X,Y,z(U)),
\end{equation}
which implies $X=Y$. When combined with the markov chain $X-(U,V)-Y$, this gives $H(X|UV)=0$. Therefore, $\Rcal_{\whpcrit} \subseteq \Scal$ follows from writing
\begin{IEEEeqnarray*}{rCl}
  R &\geq& I(X;U,V) = H(X)\\
  R_0 &\geq& I(W_x;V|U)=I(W_x;X,V|U)\geq I(W_x;X|U).
 \end{IEEEeqnarray*}
 To see $\Scal \subseteq \Rcal_{\whpcrit}$, let $V=Y=X$.
\end{IEEEproof}

Corollary~\ref{losslessW}, in turn, spawns two important results. By invoking Corollary~\ref{losslessW} with $W_x=\emptyset$, we recover Proposition~\ref{secrecyischeap} under {\sc whp}. 
\begin{cor}
\label{losslessNOcor}
Fix $P_X$ and $\pi(x,z)$. If lossless communication is required and there is no causal disclosure, then the rate-payoff region $\Rcal_{\whpcrit}$ is equal to
 \begin{equation}
 \left\{
 \begin{IEEEeqnarraybox}[][c]{rCl}
 \IEEEeqnarraymulticol{3}{l}{
 (R,R_0,\Pi): \vspace{2pt}
 }\\
 R &\geq& H(X)\\
 R_0 &\geq& 0\\
 \Pi &\leq& \min_z \Ebb\, \pi(X,z)
 \end{IEEEeqnarraybox}
\right\}.
\end{equation}
\end{cor}
If we instead consider the disclosure channel $W_x=X$, we have the following.
\begin{cor}
\label{losslessX}
Fix $P_X$ and $\pi(x,z)$. If lossless communication is required and $X^{i-1}$ is disclosed, then the rate-payoff region $\Rcal_{\whpcrit}$ is equal to
 \begin{equation}
 \label{losslessXregion}
 \bigcup_{P_{U|X}}\left\{
 \begin{IEEEeqnarraybox}[][c]{rCl}
 \IEEEeqnarraymulticol{3}{l}{
 (R,R_0,\Pi): \vspace{2pt}
 }\\
 R &\geq& H(X)\\
 R_0 &\geq& H(X|U)\\
 \Pi &\leq& \min_{z(u)} \Ebb\, \pi(X, z(U))
 \end{IEEEeqnarraybox}
\right\}.
\end{equation}
\end{cor}
\subsection{Lossless communication example}
In this section, we present a concrete example of the region in Corollary~\ref{losslessX} (causal disclosure of Node A) and compare it to the region in Corollary~\ref{losslessNOcor} (no causal disclosure). 

We first show that \eqref{losslessXregion} can be written as a linear program.   Since the constraint on $R$ is fixed by the source distribution, we focus our attention on the boundary of the $(R_0,\Pi)$ tradeoff, namely
\begin{equation}
\Pi(R_0) \triangleq \max_{\substack{P_{U|X}:\\ H(X|U)\geq R_0}}\min_{z(u)}\Ebb\,\pi(X,z(U)).
\end{equation}
Notice that this can be rewritten as
\begin{equation}
\label{tradeoffrewrite}
\Pi(R_0) = \max_{\substack{P_{U},P_{X|U}:\\ \sum_u P_U P_{X|U} = P_X\\H(X|U)\geq R_0}}\sum_{u}P_{U}(u)\min_{z}\Ebb[\pi(X,z)|U=u].
\end{equation}
If we are able to restrict the set $\{P_{X|U=u}\}_{u\in\Ucal}$ in the maximization to a finite set $\Pcal\subseteq \Delta_{\Xcal}$, then $\Pi(R_0)$ can be expressed as a linear program. Indeed, viewing the distribution $P_U$ as a vector $p\in\Rbb^{|\Pcal|}$, \eqref{tradeoffrewrite} becomes
\begin{equation}
\begin{IEEEeqnarraybox}{ll}
 \text{maximize}\quad & d^{\top} p \\
 \text{subject to} & p \geq 0\\
 & 1^{\top} p=1\\
 & Tp=P_X\\
 & h^{\top} p \leq R_0
\end{IEEEeqnarraybox}
\end{equation}
where 
\begin{itemize}
\item $T\in\Rbb^{|\Xcal|\times|\Pcal|}$ is the transition matrix whose columns are the elements of $\Pcal$.
\item The vector $d\in\Rbb^{|\Pcal|}$ has entries 
\begin{equation}
d_u=\min_z\Ebb[\pi(X,z)|U=u],\,u\in\Ucal.
\end{equation}
\item The vector $h\in\Rbb^{|\Pcal|}$ has entries 
\begin{equation}
h_u=H(X|U=u),\,u\in\Ucal.
\end{equation}
\end{itemize}
To see why there is always a choice of finite $\Pcal$ such that the rate-payoff boundary is unaffected, consider the function $d:\Delta_{\Xcal}\rightarrow \Rbb$ defined by 
\begin{equation}
d(p)=\min_{z}\Ebb[\pi(X,z)], \mbox{where } X\sim p.
\end{equation}
Observe that $d(\cdot)$ is the boundary of a convex polytope because it is the minimum of $|\Zcal|$ linear functions (and $\Zcal$ is finite). Define the set 
\begin{equation}
\Pcal=\{p\in\Delta_{\Xcal}: d(p) \mbox{ is an extreme point of } d \}
\end{equation}

  Given a set of distributions $\{P_{X|U=u}\}_{u\in\Ucal}$ that optimize \eqref{tradeoffrewrite}, we can write each element $P_{X|U=u}$ as a convex combination of the distributions in $\Pcal$ while maintaining the value of the objective. Furthermore, due to the concavity of the entropy function, the constraint on $R_0$ is still satisfied. Thus, $\Pcal$ is sufficient for the optimization. 
  
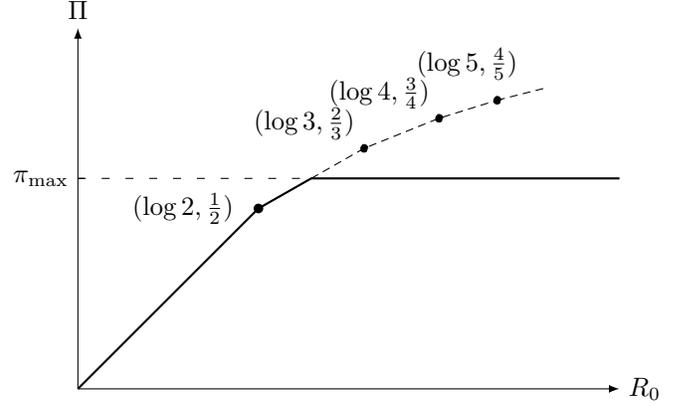
\begin{figure}
\centering
\begin{tikzpicture}[scale=2.4,
 dot/.style={draw=black,fill=black,circle,minimum size=1mm,inner sep=0pt}]
 \draw[arw] (0,0) to node[above=4mm,pos=0.3,sloped] {} (0,2) node[above] {$\Pi$};
 \draw[arw] (0,0) to node[below=2mm,midway,sloped] {} (3,0) node[right]{$R_0$};
 \draw (1pt,2*7/12) -- (0,2*7/12) node[anchor=east] {$\pi_{\max}$};

 \draw[black,densely dashed] (${ln(2)/ln(2)+ln(3)/ln(2)}*(1/2,0)+(0,1/2+2/3)$) -- (${ln(3)/ln(2)}*(1,0)+2*(0,2/3)$) node[dot] {} node[above left] {$(\log3,\frac{2}{3})$} --(${ln(4)/ln(2)}*(1,0)+2*(0,3/4)$) node[dot] {} node[above left] {$(\log4,\frac{3}{4})$} --(${ln(5)/ln(2)}*(1,0)+2*(0,4/5)$) node[dot] {} node[above=1.5mm,xshift=-4mm] {$(\log5,\frac{4}{5})$};
 \draw[black,loosely dashed] (0,2*7/12)--(${ln(2)/ln(2)+ln(3)/ln(2)}*(1/2,0)+(0,1/2+2/3)$);
 \draw[black,densely dashed] (${ln(5)/ln(2)}*(1,0)+2*(0,4/5)$) -- (${ln(6)/ln(2)}*(1,0)+2*(0,5/6)$);

 \draw[black,thick] (0,0) -- (${ln(2)/ln(2)}*(1,0)+2*(0,1/2)$) node[dot] {} node[black,left=2mm] {$(\log2,\frac{1}{2})$} -- (${ln(2)/ln(2)+ln(3)/ln(2)}*(1/2,0)+(0,1/2+2/3)$) -- (3,2*7/12);
\end{tikzpicture}
\caption{\small Illustration of Theorem~\ref{losslessexample} for a generic source $P_X$ with $1-\max_xP_X(x)=\pi_{\max}$. The solid curve, $\Pi(R_0)=\min\{\phi(R_0),\pi_{\max}\}$, is the tradeoff between rate of secret key and payoff under the assumption of causal disclosure (Corollary~\ref{losslessX}). The loosely dashed line is $\pi_{\max}$, which also corresponds to the payoff when there is no causal disclosure (Corollary~\ref{losslessNOcor}). The densely dashed curve is $\phi(R_0)$.}
\label{fig:losslesstradeoff}
\end{figure}

In the particular case that the payoff function is hamming distance (i.e., $\pi(x,z)=\bold{1}\{x\neq~\!\!z\}$), the set $\Pcal$ has a particularly convenient form:
\begin{equation}
\Pcal = \{ p\in \Delta_\Xcal: p=\mbox{Unif}(\Acal) \mbox{ for some }\Acal \subseteq \Xcal \}.
\end{equation}
This allows us to give the following simple analytical expression for $\Pi(R_0)$. The proof is given in Appendix~\ref{losslessexampleappendix}.
\begin{thm}
\label{losslessexample}
 Fix $P_X$ and let $\pi(x,z)=\bold{1}{\{x\neq z\}}$. Define the function $\phi(\cdot)$ as the linear interpolation of the points $(\log n,\frac{n-1}{n}),n\in\Nbb$.\footnote{Here $n$ does not refer to blocklength.} Also, define 
 \begin{equation}
 \pi_{\max}= 1-\max_x P_X(x). 
 \end{equation}
Then, the boundary of the rate-payoff region when lossless communication is required and $X^{i-1}$ is disclosed can be written as 
\begin{equation}
 \Pi(R_0)=\min\{\phi(R_0),\pi_{\max}\}.
\end{equation}
\end{thm}
In Fig.~\ref{fig:losslesstradeoff}, we illustrate Theorem~\ref{losslessexample} for an arbitrary source distribution.  Note that when there is \emph{no} causal disclosure and $\pi(x,z)$ is hamming distance, the payoff is given by Corollary~\ref{losslessNOcor} as
\begin{equation}
\label{losslessNOexample}
\min_z \Ebb\,\pi(X,z)=1-\max_x P_X(x)=\pi_{\max},
\end{equation}
regardless of the rate of secret key. Comparing \eqref{losslessNOexample} with $\min\{\phi(R_0),\pi_{\max}\}$ demonstrates the effect of causal disclosure (see Fig.~\ref{fig:losslesstradeoff}). In particular, we see that the assumption that the adversary does not view any of the true source bits can lead to a rather fragile guarantee of maximum secrecy. Indeed, at low rates of secret key, the gap that results from revealing the source causally is the difference between maximum secrecy and zero secrecy. This reduction in payoff is the price that is paid for increased robustness against an adversary (e.g., preventing pitfalls like those that we saw in the example of one-bit secrecy).

From Theorem~\ref{losslessexample}, we also readily see that the payoff can saturate when $R_0 < H(X)$, which shows that maximum payoff is not the same as Shannon's perfect secrecy. For example, if $P_X=\{1/4,1/4,1/2\}$, then the maximum payoff of $1/2$ occurs at $R_0=1$, but $H(X)=1.5$.
\subsection{Lossy communication}
In the previous section, the communication rate lay above $H(X)$ and did not affect the $(R_0,\Pi)$ tradeoff. However, when the requirement of lossless communication is relaxed, all three quantities interact. There are four natural special cases that are obtained by setting $W_x$ equal to $\emptyset$ or $X$ and setting $W_y$ equal to $\emptyset$ or $Y$. We denote the corresponding rate-payoff regions as $\Rcal_{\emptyset}$, $\Rcal_A$, $\Rcal_B$, and $\Rcal_{AB}$ to distinguish which nodes' actions are causally revealed.
\begin{cor}
\label{fourcases}
Fix $P_X$ and $\pi(x,y,z)$. In each of the following, the region holds under all three payoff criteria.

If there is no causal disclosure, then the rate-payoff region, $\Rcal_{\emptyset}$, is equal to
 \begin{equation}
 \label{lossyNOregion}
 \bigcup_{P_{Y|X}}\left\{
 \begin{IEEEeqnarraybox}[][c]{rCl}
 \IEEEeqnarraymulticol{3}{l}{
 (R,R_0,\Pi): \vspace{2pt}
 }\\
 R &\geq& I(X;Y)\\
 R_0 &\geq& 0\\
 \Pi &\leq& \min_{z} \Ebb\, \pi(X,Y,z)
 \end{IEEEeqnarraybox}
\right\}.
\end{equation}
If $X^{i-1}$ is disclosed, then the rate-payoff region, $\Rcal_A$, is equal to
 \begin{equation}
 \label{lossyXregion}
 \bigcup_{P_{Y,U|X}}\left\{
 \begin{IEEEeqnarraybox}[][c]{rCl}
 \IEEEeqnarraymulticol{3}{l}{
 (R,R_0,\Pi): \vspace{2pt}
 }\\
 R &\geq& I(X;Y,U)\\
 R_0 &\geq& I(X;Y|U)\\
 \Pi &\leq& \min_{z(u)} \Ebb\, \pi(X,Y,z(u))
 \end{IEEEeqnarraybox}
\right\}.
\end{equation}
If $Y^{i-1}$ is disclosed, then $\Rcal_B$ is given by directly substituting $W_x=\emptyset$ and $W_y=Y$ in \eqref{mainthmregion}. Similarly, if $(X^{i-1},Y^{i-1})$ is disclosed, then $\Rcal_{AB}$ is given by directly substituting $W_x=X$ and $W_y=Y$ in \eqref{mainthmregion}.
\end{cor}

 \begin{IEEEproof}
 Setting $(W_x,W_y)=(\emptyset,\emptyset)$ in Theorem~\ref{mainthm} gives $\Rcal_{\emptyset}$. Denote the region in \eqref{lossyNOregion} by $\Scal$. If $(R,R_0,\Pi)\in\Rcal_{\emptyset}$, then
 \begin{IEEEeqnarray}{l}
  R \geq I(X;U,V) = I(X;U,V,Y) \geq I(X;Y)\\
  \Pi \leq \min_{z(u)}\Ebb\,\pi(X,Y,z(U))\leq \min_{z}\Ebb\,\pi(X,Y,z),
 \end{IEEEeqnarray}
 which gives $\Rcal_{\emptyset}\subseteq\Scal$. To see $\Scal\subseteq\Rcal_{\emptyset}$, let $U=\emptyset$ and $V=Y$.
 
 Setting $(W_x,W_y)=(X,\emptyset)$ in Theorem~\ref{mainthm} gives $\Rcal_A$. Denote the region in \eqref{lossyXregion} by $\Tcal$. If $(R,R_0,\Pi)\in\Rcal_A$, then
\begin{IEEEeqnarray}{l}
  R \geq I(X;U,V) = I(I;U,V,Y) \geq I(X;U,Y)\\
  R_0 \geq I(X;V|U) = I(X;V,Y|U) \geq I(X;Y|U),
 \end{IEEEeqnarray}
 which gives $\Rcal_A\subseteq\Tcal$. To see $\Tcal\subseteq\Rcal_A$, let $V=Y$.
 \end{IEEEproof}

\subsection{Lossy communication examples}
In this section, we investigate concrete examples of Corollary~\ref{fourcases} by considering the payoff function
\begin{equation}
\label{lossyexamplepayoff}
\pi(x,y,z) = \bold{1}\{x=y, x\neq z\}.
\end{equation}

For this choice, the block-average payoff is the fraction of symbols in a block that Nodes A and B are able to agree on and keep hidden from Node C. 

We now present achievable regions for the cases of Corollary~\ref{fourcases} when $P_{X}\sim\text{Bern}(1/2)$ and $\pi(x,y,z)$ is given by \eqref{lossyexamplepayoff}. The region that we give for $R_{\emptyset}$ is optimal, and numerical computation suggests that the other regions are optimal as well. 
Setting $P_{Y|X}=\text{BSC}(\alpha)$, we have
\begin{equation}
\label{example0}
 \Rcal_{\emptyset} = \bigcup_{\alpha\in[0,\frac12]}\left\{
 \begin{IEEEeqnarraybox}[][c]{rCl}
 \IEEEeqnarraymulticol{3}{l}{
 (R,R_0,\Pi): \vspace{2pt}
 }\\
 R &\geq& 1-h(\alpha)\\
 R_0 &\geq& 0\\
 \Pi &\leq& \tfrac12 (1-\alpha)
 \end{IEEEeqnarraybox}
\right\}.
\end{equation}
If we let $U=\emptyset$ and $P_{Y|X}=\text{BSC}(\alpha)$, then we have
 \begin{equation}
 \label{exampleA}
 \Rcal_A \supseteq \bigcup_{\alpha\in[0,\frac12]}\left\{
 \begin{IEEEeqnarraybox}[][c]{rCl}
 \IEEEeqnarraymulticol{3}{l}{
 (R,R_0,\Pi): \vspace{2pt}
 }\\
 R &\geq& 1-h(\alpha)\\
 R_0 &\geq& 1-h(\alpha)\\
 \Pi &\leq& \tfrac12 (1-\alpha)
 \end{IEEEeqnarraybox}
\right\}.
\end{equation}
Letting $U=\emptyset$, $P_{Y|X}=\text{BSC}(\alpha)$, and $P_{V|Y}=\text{BSC}(\beta)$ gives
 \begin{equation}
 \label{exampleB}
 \Rcal_B \supseteq \bigcup_{\alpha,\beta\in[0,\frac12]}\left\{
 \begin{IEEEeqnarraybox}[][c]{rCl}
 \IEEEeqnarraymulticol{3}{l}{
 (R,R_0,\Pi): \vspace{2pt}
 }\\
 R &\geq& 1-h(\alpha)\\
 R_0 &\geq& 1-h(\beta)\\
 \Pi &\leq& \tfrac12 (1-\alpha \star \beta)
 \end{IEEEeqnarraybox}
\right\}
\end{equation}
and also
 \begin{equation}
 \Rcal_{AB} \supseteq \text{conv}\left(\bigcup_{\alpha,\beta\in[0,\frac12]}\left\{
 \begin{IEEEeqnarraybox}[][c]{rCl}
 \IEEEeqnarraymulticol{3}{l}{
 (R,R_0,\Pi): \vspace{2pt}
 }\\
 R &\geq& 1-h(\alpha)\\
 R_0 &\geq& 1+h(\alpha\star\beta)\\
 && -\> h(\alpha)-h(\beta)\\
 \Pi &\leq& \tfrac12 (1-\alpha \star \beta)
 \end{IEEEeqnarraybox}
\right\}\right).
\end{equation}
where $\alpha\star\beta=\alpha(1-\beta)+\beta(1-\alpha)$ and $\text{conv}(\cdot)$ denotes the convex hull operation. Regions \eqref{exampleA} and \eqref{exampleB} are convex as given.
\begin{figure}
\centering
\begin{subfigure}[t]{0.4\textwidth}
\includegraphics{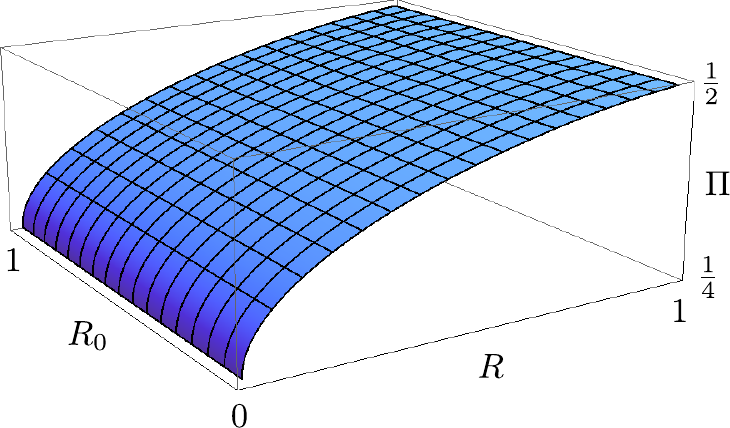}
\caption{No causal disclosure.}
\label{lossyexampleNo}
\end{subfigure}
\\
\begin{subfigure}[t]{0.4\textwidth}
\includegraphics{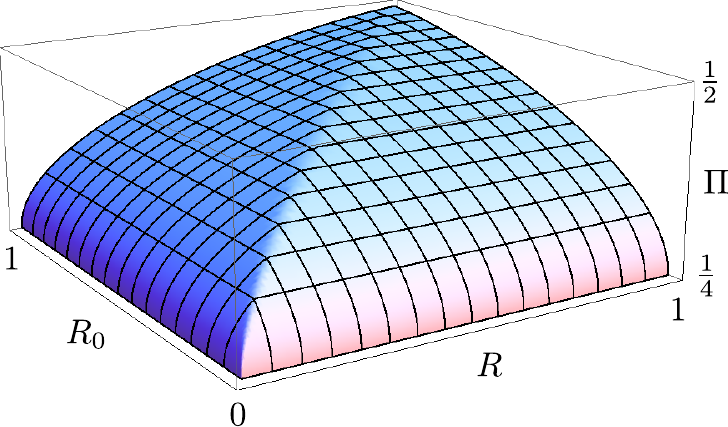}
\caption{Node A causally disclosed.}
\label{lossyexampleA}
\end{subfigure}
\\
\begin{subfigure}[t]{0.4\textwidth}
\includegraphics{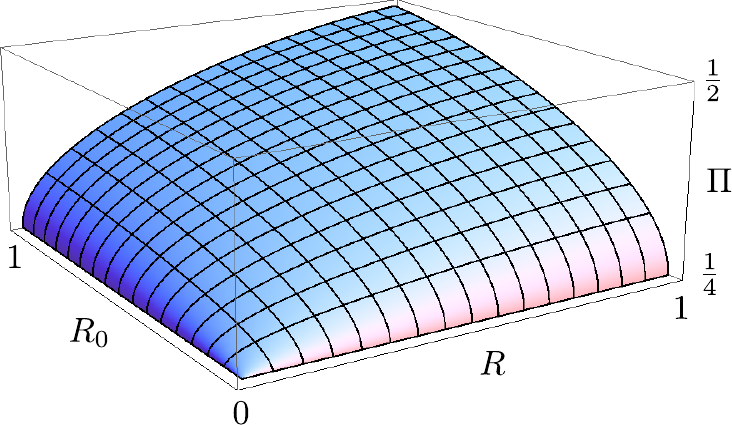}
\caption{Node B causally disclosed.}
\label{lossyexampleB}
\end{subfigure}
\\
\begin{subfigure}[t]{0.4\textwidth}
\includegraphics{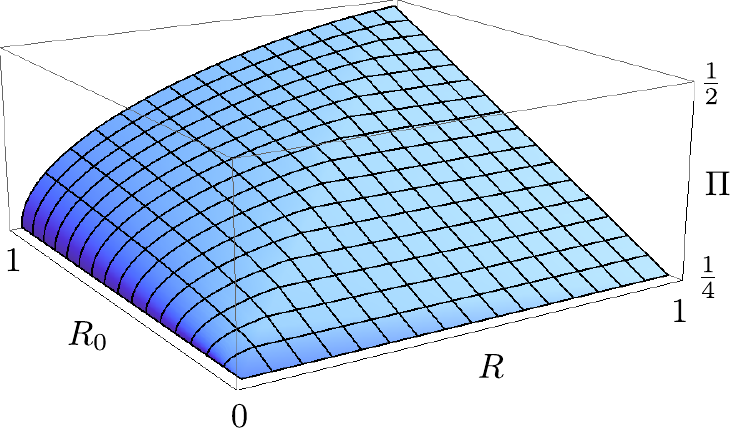}
\caption{Nodes A and B causally disclosed.}
\label{lossyexampleAB}
\end{subfigure}
\caption{\small Achievable regions of Corollary~\ref{fourcases} for $P_X\sim\text{Bern}(1/2)$ and $\pi(x,y,z)=\bold{1}\{x=y, x\neq z\}$. Numerical computation suggests that these regions are optimal.} 
\label{fig:lossyexamples}
\end{figure}

Several observations concerning the regions in Fig.~\ref{fig:lossyexamples} are in order. First, the minimum payoff is $1/4$, which occurs when there is no communication or secret key. This is achieved if Node B generates an i.i.d. sequence according to $\text{Bern}(1/2)$ and Node C produces an arbitrary sequence. The maximum payoff that can be guaranteed is $1/2$, because Node C can correctly guess $X$ with probability one-half without any information. Second, note the strict containment from top to bottom: causal access to Node A (Fig.~\ref{lossyexampleA}) is better for the adversary than access to Node B (Fig.~\ref{lossyexampleB}), and the combination (Fig.~\ref{lossyexampleAB}) is strictly better for him than Node A alone. Finally, observe the effect of having a higher secret key rate than communication rate, and vice versa. When Node A is causally revealed, the payoff is a function of $\min(R,R_0)$ and there is no advantage in having excess of either rate. However, when Node B is revealed, both $R_0>R$ and $R>R_0$ result in higher payoff than $R=R_0$.  When both nodes are revealed, an excess of secret key rate increases payoff.\footnote{These relationships are not known to be true in general.} This phenomenon is particularly surprising because it means that secret key is useful even beyond the application of a one-time-pad to the communication.

\section{Equivocation}
\label{sec:equivocation}
In this section, we show that (normalized) equivocation-based measures of secrecy become a special case of the causal disclosure framework if we choose the payoff function to be a log-loss function. Relating distortion to conditional entropy via a log-loss function was done recently in the context of certain multiterminal source coding problems \cite{Courtade2013}.

First, we remark that Theorem~\ref{mainthm} can be readily extended to include multiple distortion functions. For example, if we wanted to separately evaluate coordination and secrecy, we could use two payoff functions $\pi_1(x,y)$ and $\pi_2(x,y,z)$. In this setting, it might be more natural to refer to distortion functions than payoff functions, with the goal of minimizing the distortion between Nodes A and B while maximizing the distortion between Nodes (A,B) and Node C. Then, the rate-distortion region becomes
\begin{equation}
\label{multipledistortionregion}
 \bigcup_{W_x-X-(U,V)-Y-W_y}\left\{
 \begin{IEEEeqnarraybox}[][c]{rCl}
 \IEEEeqnarraymulticol{3}{l}{
 (R,R_0,D_1,D_2): \vspace{2pt}
 }\\
 R &\geq& I(X;U,V)\\
 R_0 &\geq& I(W_xW_y;V|U)\\
 D_1 &\geq& \Ebb\, d_1(X,Y)\\
 D_2 &\leq& \min_{z(u)}\Ebb\,d_2(X,Y,z(U))
 \end{IEEEeqnarraybox}
\right\}.
\end{equation}
Now consider $(W_x,W_y)=(X,\emptyset)$ and a distortion function $d_2:\Xcal\times\Ycal\times\Delta_{\Xcal}\rightarrow \Rbb$ defined by
\begin{equation}
\label{loglossX}
d_2(x,y,z)=\log \frac{1}{z(x)}
\end{equation}
where $z$ is a probability distribution on $\Xcal$, and $z(x)$ denotes the probability of $x\in\Xcal$ according to $z\in\Delta_{\Xcal}$. With this choice, the distortion in criterion {\sc avg} can be written as
\begin{IEEEeqnarray}{rCl}
\IEEEeqnarraymulticol{3}{l}{
\nonumber\min_{\{P_{Z_i|M,X^{i-1}}\}_{i=1}^n}\Ebb\,\frac1n \sum_{i=1}^n d_2(X_i,Y_i,Z_i)
}\\
\quad &=& \frac1n \sum_{i=1}^n \min_{P_{Z|M,X^{i-1}}}\Ebb\, d_2(X_i,Y_i,Z)\\
&=& \frac1n \sum_{i=1}^n \min_{P_{Z|M,X^{i-1}}}\Ebb\, \log\frac{1}{Z(X_i)}\\
&\stackrel{(a)}{=}& \frac1n \sum_{i=1}^n H(X_i|M,X^{i-1})\\
&=& \frac1n H(X^n|M),
\end{IEEEeqnarray}
where (a) is due to the Lemma~\ref{loglosslemma} (given below).  
Thus, for the log-loss distortion function in \eqref{loglossX}, expected adversarial distortion under an assumption of causal disclosure simply becomes normalized equivocation. 

\begin{lemma}
\label{loglosslemma}
Fix a pair of random variables $(X,Y)$ and let $\Zcal=\Delta_{\Xcal}$. Then
\begin{equation}
H(X|Y)=\min_{Z:\, X-Y-Z} \Ebb\,\log\frac{1}{Z(X)}
\end{equation}
where $z(x)$ is the probability of $x$ according to $z$.
\end{lemma}
\begin{IEEEproof}
If $X-Y-Z$, then
\begin{IEEEeqnarray}{rCl}
\IEEEeqnarraymulticol{3}{l}{
\Ebb \,\log\frac{1}{Z(X)}
}\\
\quad &=& \Ebb \,\log\frac{1}{P_{X|Y}(X|Y)}+\Ebb \,\log\frac{P_{X|Y}(X|Y)}{Z(X)}\\
&=& H(X|Y) + \sum_{y,z}P_{YZ}(y,z)D(P_{X|Y=y}||z)\\
&\geq& H(X|Y),
\end{IEEEeqnarray}
with equality if $z=P_{X|Y=y}$ for all $(y,z)$.
\end{IEEEproof}

So far, we have focused on the equivocation of $X^n$; however, one might be interested in $\frac1n H(Y^n|M)$ or $\frac1n H(X^n,Y^n|M)$, instead. In these cases, the rate-distortion-equivocation regions can again be recovered from Theorem~\ref{mainthm} (via the form in \eqref{multipledistortionregion}) by considering $(W_x,W_y)=(\emptyset,Y)$, $\Zcal=\Delta_{\Ycal}$ and \begin{equation}
\label{loglossY}
d_2(x,y,z)=\log\frac{1}{z(y)}
\end{equation}
or $(W_x,W_y)=(X,Y)$,  $\Zcal=\Delta_{\Xcal\times\Ycal}$ and 
\begin{equation}
d_2(x,y,z)=\log\frac{1}{z(x,y)}.
\end{equation}

In all three cases, the regions can be simplified (in particular, the auxiliary random variable $V$ can be eliminated). The results are given in the following theorem, part 1 of which was given by Yamamoto in \cite{Yamamoto1997}.
\begin{cor}
Fix $P_X$ and $d(x,y)$. Let $\Rcal$ denote the closure of achievable pairs $(R,R_0,D,E)$.
\begin{enumerate}[1)]
\item If the equivocation criterion is
\begin{equation}
\liminf_{n\to\infty}\frac1n H(X^n|M) \geq E,
\end{equation}
then
\begin{equation}
 \Rcal = \bigcup_{P_{Y|X}}\left\{
 \begin{IEEEeqnarraybox}[][c]{rCl}
 \IEEEeqnarraymulticol{3}{l}{
 (R,R_0,D,E): \vspace{2pt}
 }\\
 R &\geq& I(X;Y)\\
 D &\geq& \Ebb\, d(X,Y)\\
 E &\leq& H(X) - [I(X;Y)-R_0]_+
 \end{IEEEeqnarraybox}
\right\},
\end{equation}
where $[x]_+=\max\{0,x\}$.
\item If the equivocation criterion is
\begin{equation}
\liminf_{n\to\infty}\frac1n H(Y^n|M) \geq E,
\end{equation}
then
\begin{equation}
\label{equivocationYregion}
 \Rcal = \bigcup_{X-U-Y}\left\{
 \begin{IEEEeqnarraybox}[][c]{rCl}
 \IEEEeqnarraymulticol{3}{l}{
 (R,R_0,D,E): \vspace{2pt}
 }\\
 R &\geq& I(X;U)\\
 D &\geq& \Ebb\, d(X,Y)\\
 E &\leq& H(Y) - [I(Y;U)-R_0]_+
 \end{IEEEeqnarraybox}
\right\}.
\end{equation}
\item If the equivocation criterion is
\begin{equation}
\liminf_{n\to\infty}\frac1n H(X^n,Y^n|M) \geq E,
\end{equation}
then
\begin{equation}
 \Rcal = \bigcup_{X-U-Y}\left\{
 \begin{IEEEeqnarraybox}[][c]{rCl}
 \IEEEeqnarraymulticol{3}{l}{
 (R,R_0,D,E): \vspace{2pt}
 }\\
 R &\geq& I(X;U)\\
 D &\geq& \Ebb
 \, d(X,Y)\\
 E &\leq& H(X,Y) - [I(X,Y;U)-R_0]_+
 \end{IEEEeqnarraybox}
\right\}.
\end{equation}
\end{enumerate}
\end{cor}

\begin{IEEEproof}
We only prove part 2, as parts 1 and 3 follow similar arguments. First, fix $d_2(x,y,z)$ according to \eqref{loglossY}. Then, by Lemma~\ref{loglosslemma},
\begin{equation}
\min_{z(u)}\Ebb\,d_2(X,Y,z(U)) = H(Y|U).
\end{equation}
From the discussion above, it is clear that $\Rcal$ is characterized by setting $(W_x,W_y)=(\emptyset,Y)$ in \eqref{multipledistortionregion}, yielding
\begin{equation}
 \Rcal=\bigcup_{X-(U,V)-Y}\left\{
 \begin{IEEEeqnarraybox}[][c]{rCl}
 \IEEEeqnarraymulticol{3}{l}{
 (R,R_0,D,E): \vspace{2pt}
 }\\
  R &\geq& I(X;U,V)\\
 R_0 &\geq& I(Y;V|U)\\
 D &\geq& \Ebb\, d(X,Y)\\
 E &\leq& H(Y|U) 
 \end{IEEEeqnarraybox}
\right\}.
\end{equation}
Denote the region in \eqref{equivocationYregion} by $\Scal$. To see $\Rcal\subseteq\Scal$, first consider $(R,R_0,D,E)\in\Rcal$. Defining $U'\triangleq (U,V)$, we have
\begin{IEEEeqnarray}{rCl}
  R &\geq& I(X;U,V) = I(X;U')\\
  E &\leq& H(Y|U)=H(Y|U,V)+I(Y;V|U)\\
  &\leq& H(Y|U')+R_0\\
  E &\leq& H(Y),
 \end{IEEEeqnarray}
which implies $(R,R_0,D,E)\in \Scal$. To see $\Scal\subseteq\Rcal$, let $(R,R_0,D,E)\in\Scal$. Define $V'\triangleq U$ and find a random variable $U'$ such that $U'-U-(X,Y)$ form a markov chain and 
\begin{equation}
\label{equiv_step_aux}
H(Y|U')=H(Y) - [I(Y;U)-R_0]_+
\end{equation}
This is always possible because the right-hand side of \eqref{equiv_step_aux} lies in the interval $[H(Y|U),H(Y)]$. Then, we can write
\begin{IEEEeqnarray}{rCl}
  R &\geq& I(X;U) =I(X;U',V')\\
  R_0 &\geq& H(Y|U')-H(Y|U)=I(Y;V'|U')\\
  E &\leq& H(Y|U'),
 \end{IEEEeqnarray}
which implies $(R,R_0,D,E)\in \Rcal$. Thus, $\Rcal=\Scal$.
\end{IEEEproof}
\section{Achievability proof}
\label{sec:achievability}
\subsection{Soft covering lemma}
The primary tool used in the achievability proof of Theorem~\ref{mainthm} is a so-called ``soft covering lemma", a known result concerning the approximation of the output distribution of a channel.\footnote{The name ``soft covering lemma" was given in \cite{Cuff2013}. The same lemma has also been referred to as the ``resolvability lemma" and ``cloud-mixing lemma".} Various forms of the lemma have appeared in \cite{Wyner1975common} and \cite{Han1993} and related notions from the perspective of random binning can be found in \cite{Yassaee2014}. Several generalizations of the lemma (including a one-shot version) can be found in \cite{Cuff2013}.

 In brief, the most basic version of the soft covering lemma is as follows. Fix a joint distribution $P_{X,U}$. First, generate a random codebook of $2^{nR}$ independent codewords, each drawn according to $\prod_{i=1}^n P_U(u_i)$. Select a codeword, uniformly at random, as the input to a memoryless channel $\prod_{i=1}^nP_{X|U}(x_i|u_i)$. The lemma states that if $R>I(X;U)$, then the distribution of the channel output $X^n$ converges to $\prod_{i=1}^n P_X(x_i)$ in expected total variation distance, where the expectation is with respect to the random codebook.
 
A generalization of the soft covering lemma, presented shortly, will prove essential to the payoff analysis. Once we define a code by pairing a random codebook with a particular stochastic encoder and decoder, the soft covering lemma can be used to approximate the joint statistics of the system (i.e., the joint distribution on $(X^n,M,K,Y^n,W^n)$ that is induced by the code) by an ``idealized" distribution that has desirable properties. Having a tractable approximation of the joint distribution of the system is important because an adversary's optimal strategy is dictated by a posterior distribution. For example, if an adversary tries to estimate the $i$th source symbol $X_i$ based on his observations of causal disclosure $X^{i-1}$ and the public message $M$, his optimal strategy is entirely determined by the posterior distribution of $X_i$ given $(X^{i-1},M)$. The approximating distribution that the soft covering lemma guarantees will provide a clear understanding of that posterior distribution and lead to a manageable payoff analysis.

Although the distribution approximation in the soft covering lemma holds for normalized and unnormalized divergence, we use the total variation version found in \cite{Cuff2013} and \cite{Han1993} because of the following properties that total variation enjoys.

Given two probability measures $P$ and $Q$ with common alphabet $\Xcal$, the total variation distance between $P$ and $Q$ is defined by
\begin{equation}
\lVert P - Q \rVert = \sup_{A\in\mathcal{F}}|P(A)-Q(A)|,
\end{equation}
where $\mathcal{F}$ is the sigma algebra of the common alphabet.

\begin{property}
Total variation distance satisfies the following.
\begin{enumerate}[(a)]
\item If the support of $P$ and $Q$ is a countable set $\Xcal$, then
\begin{equation}
\lVert P - Q \rVert = \frac12 \sum_{x\in\Xcal} \Big |P(\{x\})-Q(\{x\})\Big |.
\end{equation}
\item Let $\eps>0$ and let $f(x)$ be a function with bounded range of width $b>0$. Then
\begin{equation}
\lVert P-Q \rVert < \eps \:\Longrightarrow\: \big| \Ebb_Pf(X) - \Ebb_Qf(X) \big | < \eps b,
\end{equation}
where $\Ebb_{P}$ indicates that the expectation is taken with respect to the distribution $P$.
\item For any $P$, $Q$, and $\Phi$,
\begin{equation}
\lVert P - Q \rVert \leq \lVert P - \Phi \rVert + \lVert \Phi - Q \rVert.
\end{equation}
\item Let $P_{X}P_{Y|X}$ and $Q_XP_{Y|X}$ be two joint distributions with common channel $P_{Y|X}$. Then
\begin{equation}
\lVert P_XP_{Y|X} - Q_X P_{Y|X} \rVert = \lVert P_X - Q_X \rVert.
\end{equation}
\item Let $P_X$ and $Q_X$ be marginal distributions of $P_{XY}$ and $Q_{XY}$. Then
\begin{equation}
\lVert P_X - Q_X \rVert \leq \lVert P_{XY} - Q_{XY} \rVert.
\end{equation}
\end{enumerate}
\end{property}
We require the following generalization of the soft covering lemma.
\begin{defn}
Let $\{P_{X^n,Y^n}\}_{n=1}^{\infty}$ be a sequence of joint distributions. The sup-information rate of this sequence is defined as
\begin{equation}
\ovr{I}(X;Y)\triangleq \limsup_{n\to\infty} \frac1n i_{P_{X^n,Y^n}}(X^n;Y^n),
\end{equation}¥
where
\begin{equation}
 \limsup_{n\to\infty} W_n \triangleq \inf\{\tau: \Pbb[W_n>\tau]\to 0\}
\end{equation}¥
and
\begin{equation}
i_{P_{X,Y}}(a;b)\triangleq \log \frac{P_{X,Y}(a,b)}{P_X(a)P_Y(b)}.
\end{equation}¥
The function $i_{P_{X,Y}}(a;b)$ is called the information density.
\end{defn}

\begin{lemma}[{\cite[Corollary VII.4]{Cuff2013},\cite{Han1993}}]
\label{generalcloudmixing}
Let  $\{P_{X^n,Y^n}\}_{n=1}^{\infty}$ be a sequence of joint distributions. Let $\Ccal^{(n)}$ be a random codebook of $2^{nR}$ sequences in $\Xcal^n$, each drawn independently according to $P_{X^n}$ and indexed by $m\in[2^{nR}]$. Let $Q_{Y^n}$ denote the output distribution of the channel when the input is selected from $\Ccal^{(n)}$ uniformly at random; that is,
\begin{equation}
Q_{Y^n}(y^n)=2^{-nR} \sum_{m\in[2^{nR}]} P_{Y^n|X^n}(y^n|X^n(m)).
\end{equation}
If $R>\ovr{I}(X;Y)$, then
\begin{equation}
\label{scl_adv_result}
\lim_{n\to\infty} \Ebb_{\Ccal^{(n)}} \lVert Q_{Y^n} - P_{Y^n}\rVert = 0,
\end{equation}
where $\Ebb_{\Ccal^{(n)}}$ indicates that the expectation is with respect to the random codebook.\footnote{Because the codebook is random, the output distribution $Q_{Y^n}$ is a random variable taking values on $\Delta_{\Ycal^n}$. One way to notate this is through the use of conditional distributions (i.e., write $Q_{Y^n|\Ccal^{(n)}}$), but we choose to suppress such notation in order to simplify the presentation.} Furthermore, the convergence in \eqref{scl_adv_result} occurs exponentially quickly with $n$ if the distribution $P_{X^nY^n}$ is memoryless.
\end{lemma}

We now begin the achievability proof of Theorem~\ref{mainthm} by specifying the random codebook, stochastic encoder, and stochastic decoder.

\subsection{Design of codebook, encoder, and decoder}
In the statement of Theorem~\ref{mainthm}, we are given disclosure channels $P_{W_x|X}$ and $P_{W_y|Y}$. For simplicity, we treat the channels as a single channel\footnote{The decomposition of the channel $P_{W|XY}$ into two channels does not play a role in the achievability proof. The reason Theorem~\ref{mainthm} does not feature a generic channel $P_{W|XY}$ is that a matching converse proof has not been supplied.} defined by
\begin{equation}
P_{W|XY} \triangleq P_{W_x|X}P_{W_y|Y}.
\end{equation}
Thus, we denote $(W_x^n,W_y^n)$ by $W^n$ and the causal disclosure by $W^{i-1}$. The memoryless channel from $(X^n,Y^n)$ to $W^n$ is denoted by
\begin{equation}
P_{W^n|X^nY^n} \triangleq \prod_{i=1}^n P_{W|XY}.
\end{equation}

Given a source distribution $P_X$ and a disclosure channel $P_{W|XY}$, fix a distribution
\begin{equation}
\label{p_factor}
P_{XUVYW} = P_{X}P_{UV|X}P_{Y|UV}P_{W|XY}.
\end{equation}
Note that this distribution satisfies the markov chain $X-(U,V)-Y$. Fix a communication rate $R>I(X;U,V)$ and a secret key rate $R_0>I(W;V|U)$.

\noindent\emph{Random codebook:} Generate a random superposition codebook in the following manner. First, generate a codebook $\Ccal_U^{(n)}$ of $2^{nR}$ codewords from $\Ucal^n$ i.i.d.\ according to $\prod_{i=1}^n P_{U}$. These codewords are indexed by $m\in[2^{nR}]$. Then, for each codeword $U^n(m) \in \Ccal^{(n)}_U$, generate a codebook $\Ccal_V^{(n)}(m)$ of $2^{nR_0}$ codewords from $\Vcal^n$ i.i.d.\ according to $\prod_{i=1}^n P_{V|U=U_i(m)}$. These codewords are indexed by $(m,k),k\in[2^{nR_0}]$. Thus, we have 
\begin{equation}
\Ccal_{U}^{(n)} = (U^n(1),\ldots,U^n(m),\ldots,U^n(2^{nR}))
\end{equation}
and
\begin{equation}
\Ccal_V^{(n)}(m) = (V^n(m,1),\ldots,V^n(m,k),\ldots,V^n(m,2^{nR_0})).
\end{equation}
We refer to the entire superposition codebook as $\Ccal^{(n)}$.

\noindent \emph{Likelihood encoder:} For a fixed superposition codebook, the encoder is a stochastic likelihood encoder defined by
\begin{equation}
\label{ll_defn}
P_{M|X^nK}(m|x^n,k) \propto \prod_{i=1}^n P_{X|U,V}(x_i | u_i(m),v_i(m,k)),
\end{equation}
where $\propto$ indicates that an appropriate normalization factor is required to make $P_{M|X^nK}$ a valid conditional probability distribution. Eqn.~\eqref{ll_defn} says that the probability of $(x^n,k)$ being mapped to the index $m$ is proportional to the probability that $x^n$ is the output of the memoryless ``test channel" $P_{X|UV}$ with input $(u^n(m),v^n(m,k))$. The reason for this choice of encoder will become clear shortly.

\noindent \emph{Decoder:} The decoder is stochastic and is defined by
\begin{equation}
P_{Y^n | MK}(y^n | m,k) \triangleq \prod_{i=1}^n P_{Y|UV}(y_i | u_i(m),v_i(m,k)).
\end{equation}

The random codebook, likelihood encoder, and decoder comprise the code and induce a joint distribution on the system that is given by
\begin{equation}
P_{X^nMKY^nW^n} = P_{X^n}P_KP_{M|X^nK}P_{Y^n|MK}P_{W^n|X^nY^n},
\end{equation}
where $P_{X^n}$ is i.i.d.\ according to $P_X$, and $P_K$ is uniform over $[2^{nR_0}]$.

\subsection{The approximating distribution $Q$ and its property}
\begin{figure*}
\centering
\begin{tikzpicture}[node distance=2cm]
 \node (dummy) [coordinate] {};
 \node (src1)   [rectangle,minimum height = 0.2cm] at ([yshift=5mm] dummy) {$M\sim\text{Unif}[2^{nR}]$};
 \node (src2)   [rectangle,minimum height = 0.2cm] at ([yshift=-5mm] dummy) {$K\sim\text{Unif}[2^{nR_0}]$};
 \node (enc1)   [node,minimum width=12mm,right=7mm of src1] {$\Ccal_U^{(n)}$};
 \node (enc2)   [node,minimum width=12mm,right=2.3cm of src2] {$\Ccal_V^{(n)}$};
 \node (split1) [coordinate] at (src1 -| enc2.center) {};
 \node (ch)     [node,minimum width=10mm,minimum height=1.5cm,right=7.3cm of dummy] {$P_{XY|UV}$};

 \node[node] (wch) [right=2cm of ch] {$P_{W|XY}$};
 \node[rectangle] (out) [right=7mm of wch] {$W^n$};

 \draw[arw] (src1) to (enc1);
 \draw[arw] (src2) to (enc2);
 \draw (enc1) to node [pos=1,above=-1mm] {$U^n(M)$} (split1);
 \draw[arw] (split1) to (src1 -| ch.west);
 \draw[arw] (split1) to (enc2);
 \draw[arw] (enc2) to node [midway,above=-1mm] {$V^n(M,K)$} (src2 -| ch.west);
 \draw[arw] (ch) to node[midway,above=-1mm] {$(X^n,Y^n)$} (wch);
 \draw[arw] (wch) to (out);
\end{tikzpicture}
\caption{\small Process that defines $Q_{X^nMKY^nW^n}$. The pair $(M,K)$ indexes a pair of codewords $(U^n(M),V^n(M,K))$ in the superposition random codebook. The codeword pair is passed through a memoryless channel $P_{XY|UV}=P_{X|UV}P_{Y|UV}$ to get $(X^n,Y^n)$. Then $(X^n,Y^n)$ is passed through a memoryless channel $P_{W|XY}$ to get $W^n$. }
\label{fig:Qdefn}
\end{figure*}
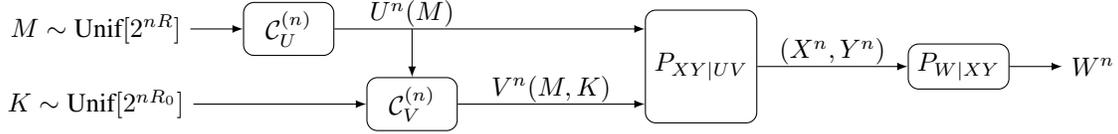

We now use the soft covering lemma (Lemma~\ref{generalcloudmixing}) to yield an approximation to the system-induced distribution $P_{X^nMKY^nW^n}$. The idealized distribution that we are concerned with is described by Fig.~\ref{fig:Qdefn} and defined explicitly as
\begin{equation}
Q_{X^nMKY^nW^n} \triangleq Q_{X^nMK}P_{Y^n|MK}P_{W^n|X^nY^n},
\end{equation}
where $Q_{X^nMK}$ is given by
\begin{equation}
Q(x^n,m,k) \triangleq 2^{n(R+R_0)} \prod_{i=1}^n P_{X|UV}(x_i|U_i(m),V_i(m,k)).
\end{equation}
Observe that the definitions of $P_{Y^n|MK}$ and $P_{W^n|X^nY^n}$, combined with the factorization of $P_{XUVYW}$ in \eqref{p_factor}, allow us to write $Q$ as
\begin{IEEEeqnarray}{l}
\nonumber Q(x^n,m,k,y^n,w^n)\\
\vspace{-8pt} \nonumber \quad = 2^{-n(R+R_0)} \prod_{i=1}^n P_{XYW|UV}(x_i,y_i,w_i|U_i(m),V_i(m,k)),\\
 ~
\end{IEEEeqnarray}
which corresponds to the process depicted in Fig.~\ref{fig:Qdefn}.

The stochastic likelihood encoder was defined intentionally so that $Q_{M|X^nK}=P_{M|X^nK}$. In fact, the only difference between $P$ and $Q$ lies in the marginal distribution of $(X^n,K)$. Indeed, notice that we can write
\begin{IEEEeqnarray}{rCl}
\IEEEeqnarraymulticol{3}{l}{
\nonumber Q_{X^nMKY^nW^n} 
}\\
\quad &\triangleq& Q_{X^nMK}P_{Y^n|MK}P_{W^n|X^nY^n}\\
&=& Q_{X^n K} P_{M | X^n K} P_{Y^n | MK}P_{W^n|X^nY^n}\\
\label{Q_difference} &=& Q_{X^n K} P_{MY^nW^n | X^n K}.
\end{IEEEeqnarray}
Therefore, we can show that $P_{X^nMKY^nW^n} \approx Q_{X^nMKY^nW^n}$ by demonstrating that $P_{X^nK}~\approx~Q_{X^nK}$. This is accomplished using the soft covering lemma.

\begin{lemma}
\label{lemma:PapproxQ}
If $R > I(X;U,V)$, then
\begin{equation}
\lim_{n\to\infty} \Ebb_{\Ccal^{(n)}}\, \big\lVert  P_{X^nMKY^nW^n} - Q_{X^nMKY^nW^n}  \big\rVert = 0.
\end{equation}
\end{lemma}
\begin{IEEEproof}[Proof of Lemma~\ref{lemma:PapproxQ}]
\begin{IEEEeqnarray}{rCl}
\IEEEeqnarraymulticol{3}{l}{\nonumber
\Ebb_{\Ccal^{(n)}}\, \big\lVert  P_{X^nMKY^nW^n} - Q_{X^nMKY^nW^n}  \big\rVert
}\\
\quad&\stackrel{(a)}{=}& \Ebb_{\Ccal^{(n)}}\, \big\lVert  P_{X^nK} - Q_{X^nK}  \big\rVert \\
&=& \Ebb_{\Ccal^{(n)}}\, \big\lVert  P_{X^n}P_{K} - Q_{X^n|K}P_{K}  \big\rVert\\
&\stackrel{(b)}{=}& 2^{-nR_0} \sum_{k=1}^{2^{nR_0}} \Ebb_{\Ccal^{(n)}}\, \big\lVert  P_{X^n} - Q_{X^n|K=k}  \big\rVert \\
&=& \Ebb_{\Ccal^{(n)}}\, \big\lVert  P_{X^n} - Q_{X^n|K=1}  \big\rVert\\
&\stackrel{(c)}{=}& o(1).
\end{IEEEeqnarray}

The justification for the steps is as follows:
\begin{enumerate}[(a)]
\item Eqn. \eqref{Q_difference} and Property 2d of total variation.
\item Property 2a of total variation.
\item $R>I(X;U,V)$ and the soft covering lemma (Lemma~\ref{generalcloudmixing}). Notice that $P_{X^n}$ is i.i.d.\ according to $P_X$ and $Q_{X^n|K=1}$ is the output distribution of the memoryless channel $P_{X|UV}$ acting on a (sub)codebook of size $2^{nR}$.
\end{enumerate}
\end{IEEEproof}
Approximating $P$ by $Q$ will allow us to analyze the payoff as if $Q$ governs the joint statistics of the system. If the rate of secret key is large enough, the structure of $Q$ will allow us to argue that the causal disclosure $W^{i-1}$ is actually useless to the eavesdropper and that his best strategy for estimating $(X_i,Y_i)$ is based solely on $U^n(M)$. The crucial property of $Q$ that enables this argument is given in the following lemma. The proof, which relies on the soft covering lemma, is relegated to  Appendix~\ref{cloudmixingappendix}. 
\begin{lemma}
\label{lemma:q_prop}
If $R_0 > I(W;V|U)$, there exists $\alpha\in(0,1]$ such that
\begin{equation}
\label{q_prop}
\lim_{n\to\infty}\Ebb_{\Ccal^{(n)}}\, \Big\lVert  Q_{MW^nX_{\Bcal}Y_{\Bcal}} - \widehat{Q}_{MW^nX_{\Bcal}Y_{\Bcal}}   \Big\rVert = 0,
\end{equation}
where
\begin{IEEEeqnarray}{l}
\nonumber \widehat{Q}_{MW^nX_{\Bcal}Y_{\Bcal}}\\
\vspace{-8pt} \nonumber \quad \triangleq Q_{M} \cdot \Big( \prod_{i=1}^n P_{W|U=U_i(M)} \Big) \cdot \Big(\prod_{i\in\Bcal} P_{XY|W,U=U_i(M)} \Big)\\
~
\end{IEEEeqnarray}
and $\Bcal$ is any subset of $[n]$ of size $|\Bcal| \leq \lfloor \alpha n \rfloor$.
\end{lemma}
To see the significance of $\widehat{Q}$, first consider $\Bcal=\emptyset$ and $W=(X,Y)$, so that
\begin{equation}
\label{artificialchannel}
\widehat{Q}_{MX^nY^n}(m,x^n,y^n)=2^{-nR}\prod_{i=1}^n P_{XY|U}(x_i,y_i|U_i(m)).
\end{equation}
Recall that $W=(X,Y)$ implies direct causal disclosure of Nodes A and B; that is, the adversary has access to $(M,X^{i-1},Y^{i-1})$ at step $i$. From \eqref{artificialchannel}, we see that $\widehat{Q}_{X^nY^n|M}$ is a memoryless channel from the codeword $U^n(M)$ to the pair $(X^n,Y^n)$. In particular, this implies
\begin{equation}
(X_i,Y_i)-U_i(M)-(M,X^{i-1},Y^{i-1}),\,\forall i\in[n].
\end{equation}
Therefore, the adversary's best estimate of $(X_i,Y_i)$ only depends on $U_i(M)$ and is not improved by the causal disclosure. We have essentially created an artificial noisy channel from the intercepted codeword $U^n(M)$ to the pair $(X^n,Y^n)$, a property which not only greatly simplifies the payoff analysis, but is interesting independent of the causal disclosure problem. We discuss this effect and some of its implications after completing the achievability proof.

For general $W$, consider $\Bcal=\{i\}$. In this case, Lemma~\ref{lemma:q_prop} demonstrates that $Q$ approximately satisfies the markov chain
\begin{equation}
\label{key_chain}
(X_i,Y_i)-U_i(M)-(M,W^{i-1}),
\end{equation}
and again we see that adversary's estimate of $(X_i,Y_i)$ only depends on $U_i(M)$ and is not improved by the causal disclosure. However, it turns out that the property in \eqref{key_chain} is not quite strong enough for the analysis of the {\sc whp} payoff criterion, which is why Lemma~\ref{lemma:q_prop} is concerned with sub-blocks $(X_{\Bcal},Y_{\Bcal})$ of size linearly increasing with $n$.

\subsection{Analysis of the {\sc min} payoff criterion}
We first combine Lemmas~\ref{lemma:PapproxQ} and~\ref{lemma:q_prop} to demonstrate the existence of a codebook that ensures certain distribution approximations hold simultaneously for all $i\in[n]$.
\begin{lemma}
\label{min_lemma}
There exists a sequence of codebooks such that
\begin{equation}
\lim_{n\to\infty} \max_{i\in[n]}\,\lVert P_{MW^n X_iY_i} - \widehat{Q}_{MW^nX_i Y_i}  \rVert =0
\end{equation}
and
\begin{equation}
\lim_{n\to\infty} \max_{i\in[n]}\, \lVert \widehat{Q}_{u_i(M)} - P_U \rVert = 0.
\end{equation}
\end{lemma}
\begin{IEEEproof}[Proof of Lemma~\ref{min_lemma}]
First, for all $i\in[n]$ we have
\begin{IEEEeqnarray}{rCl}
\IEEEeqnarraymulticol{3}{l}{\nonumber
\Ebb_{\Ccal^{(n)}}\,\lVert P_{MW^n X_iY_i} - \widehat{Q}_{MW^n X_iY_i}  \rVert
}\\
\nonumber \quad &\stackrel{(a)}{\leq} & \Ebb_{\Ccal^{(n)}}\,\lVert P_{MW^n X_iY_i} - Q_{MW^n X_iY_i}  \rVert \\
&& +\> \Ebb_{\Ccal^{(n)}}\,\lVert Q_{MW^n X_iY_i} - \widehat{Q}_{MW^n X_iY_i} \rVert\\
\nonumber &\stackrel{(b)}{\leq}& \Ebb_{\Ccal^{(n)}}\,\lVert P_{X^n MKY^nW^n} - Q_{X^n MKY^nW^n}  \rVert \\
&& +\> \Ebb_{\Ccal^{(n)}}\,\lVert Q_{MW^n X_iY_i} - \widehat{Q}_{MW^n X_iY_i} \rVert\\
\label{exponential1}&\stackrel{(c)}{=}& O(e^{-\gamma n}) 
\end{IEEEeqnarray}
for some $\gamma > 0$. Steps (a) and (b) use Properties 2c and 2e of total variation distance, respectively. Step (c) follows from Lemmas~\ref{lemma:PapproxQ} and~\ref{lemma:q_prop}, and the fact that the convergence in the soft covering lemma occurs exponentially quickly with $n$.

Next, we invoke Lemma~\ref{generalcloudmixing} to show that, for all $i\in[n]$,
\begin{equation}
\label{exponential2}
\Ebb_{\Ccal^{(n)}} \lVert  \widehat{Q}_{U_i(M)} - P_U \rVert = O(e^{-\beta n}),
\end{equation}
for some $\beta > 0$. The soft covering lemma applies because:
\begin{itemize}
\item $\widehat{Q}_{U_i(M)}$ is the output distribution of the identity channel acting on a ``codebook" of $2^{nR}$ ``codewords" generated i.i.d.\ according to $P_U$ -- the ``codebook" consists of $(U_i(1),\ldots,U_i(2^{nR}))$. Furthermore $P_U$ is the output distribution when the input distribution is $P_U$, because the channel is the identity channel.
\item The rate requirement is trivially satisfied because $R > 0$ and
\begin{equation}
\limsup_{n\to\infty} \frac1n i_{P_U}(U_i;U_i) \leq \lim_{n\to\infty} \frac1n \log |\Ucal| = 0.
\end{equation}
\end{itemize}
Combining \eqref{exponential1} and \eqref{exponential2}, we can write
\begin{IEEEeqnarray}{l}
\nonumber \lim_{n\to\infty}\Ebb_{\Ccal^{(n)}}\,\bigg[ \sum_{i=1}^n \lVert P_{X^n Y_i M} - \widehat{Q}_{X^n Y_i M} \rVert \\
\qquad \qquad \quad +\> \sum_{i=1}^n \lVert \widehat{Q}_{U_i(M)} - P_U \rVert \bigg]=0.
\end{IEEEeqnarray}
It is straightforward to verify that this fact implies the statement of the lemma.
\end{IEEEproof}

With Lemma~\ref{min_lemma} in hand, we proceed with the analysis of the {\sc min} payoff criterion. Let $\Pi \leq \min_{z(u)} \Ebb\,\pi(X,Y,z(U))$. For all $i\in[n]$, we have
\begin{IEEEeqnarray}{rCl}
\IEEEeqnarraymulticol{3}{l}{
\nonumber \min_{z(m,w^{i-1})} \Ebb_P\,\pi(X_i,Y_i,z(M,W^{i-1}))
}\\
\quad &\stackrel{(a)}{=}& \min_{z(m,w^{i-1})} \Ebb_{\widehat{Q}}\,\pi(X_i,Y_i,z(M,W^{i-1})) - o(1)\:\\
&\stackrel{(b)}{=}& \min_{z(u)} \Ebb_{\widehat{Q}}\,\pi(X_i,Y_i,z(u_i(M))) - o(1)\\
&\stackrel{(c)}{=}& \min_{z(u)} \Ebb\,\pi(X,Y,z(U)) - o(1)\\
&\geq& \Pi - o(1).
\end{IEEEeqnarray}
Step (a) uses the first part of Lemma~\ref{min_lemma} along with Property 2b of total variation. Step (b) follows from Lemma~\ref{suffstat} because under $\widehat{Q}_{MW^nX_iY_i}$, the following markov chain holds:
\begin{equation}
(X_i,Y_i)-u_i(M)-(M,W^{i-1}).
\end{equation}
Step (c) is due to the second part of Lemma~\ref{min_lemma} and Property 2b of total variation. This completes the analysis of the {\sc min} payoff criterion. 

\subsection{Analysis of the {\sc whp} payoff criterion}
Without loss of generality, we restrict attention to those distributions $P_{UVXYW}$ that satisfy
\begin{equation}
\label{noinfinity}
P_{XY}(x,y)>0 \:\Longrightarrow\: \pi(x,y,z)>-\infty\:,\forall x,y,z.
\end{equation}
Otherwise, $\min_z\Ebb\,\pi(X,Y,z)=-\infty$ and the region in Theorem~\ref{mainthm} is trivial.

The analysis will take place over sub-blocks of length $k=\lfloor\alpha n\rfloor$ rather than over the full block. For ease of presentation, we assume that $\lfloor\alpha n\rfloor=\alpha n$ and that $k$ divides $n$ evenly; the analysis is readily adjusted when this is not the case. We first fix some notation for handling sub-blocks. Denote the indices of the $j$th sub-block by the set $\Bcal_{(j)}$:
\begin{equation}
\Bcal_{(j)} = \{jk,jk+1,\ldots,(j+1)k-1\},\quad j\in[1/\alpha].
\end{equation}
Furthermore, denote the first $t$ indices of sub-block $\Bcal_{(j)}$ by $\Bcal_{(j)}^{t}$; for example, $\Bcal^1_{(j)}=j$ and $\Bcal_{(j)}^k = \Bcal_{(j)}$. Some more notation: denote the adversary's optimal reconstruction sequence by $\{Z^*_i\}_{i=1}^n$ and, for brevity, define
\begin{equation}
\rho \triangleq \min_{z(u)} \Ebb\, \pi(X,Y,z(U)).
\end{equation}

Let $\Pi < \rho$ and $\eps = \rho - \Pi$. To prove achievability under the {\sc whp} criterion, we claim that it is enough to show that, for all $j\in[1/\alpha]$,
\begin{equation}
\label{p2toshow}
\lim_{k\to\infty}\Ebb_{\Ccal^{(n)}}\,\Pbb_{\widehat{Q}}\Big[\frac1k \sum_{i\in \Bcal_{(j)}} \pi(X_i,Y_i,Z_i^*)< \rho-\eps\Big]=0,
\end{equation}
where $\widehat{Q}_{MW^nX_{\Bcal(j)}Y_{\Bcal(j)}}$ is given in Lemma~\ref{lemma:q_prop}. Indeed, if this is true, then we can write
\begin{IEEEeqnarray}{rCl}
\IEEEeqnarraymulticol{3}{l}{\nonumber
\Ebb_{\Ccal^{(n)}}\,\Pbb\Big[\frac1n \sum_{i=1}^n \pi(X_i,Y_i,Z_i^*) \geq \Pi -\eps\Big]
}\\
&\geq& \Ebb_{\Ccal^{(n)}}\,\Pbb\Big[\frac1n \sum_{i=1}^n \pi(X_i,Y_i,Z_i^*) \geq \rho-\eps\Big]\\
&\stackrel{}{\geq}& \Ebb_{\Ccal^{(n)}}\,\Pbb\Big[\bigcap_j \Big\{\frac1k \sum_{i\in \Bcal_{(j)}} \pi(X_i,Y_i,Z_i^*)\geq \rho-\eps\Big\}\Big]\\
&=& 1-  \Ebb_{\Ccal^{(n)}}\,\Pbb\Big[\bigcup_j \Big\{\frac1k \sum_{i\in \Bcal_{(j)}} \pi(X_i,Y_i,Z_i^*)< \rho-\eps\Big\}\Big]\\
&\stackrel{(a)}{\geq}& 1-\sum_{j=1}^{1/\alpha} \Ebb_{\Ccal^{(n)}}\,\Pbb\Big[\frac1k \sum_{i\in \Bcal_{(j)}} \pi(X_i,Y_i,Z_i^*)< \rho-\eps\Big]\\
\vspace{-10pt} \nonumber &\stackrel{(b)}{=}& 1-\sum_{j=1}^{1/\alpha} \Ebb_{\Ccal^{(n)}}\,\Pbb_{\widehat{Q}}\Big[\frac1k \sum_{i\in \Bcal_{(j)}} \pi(X_i,Y_i,Z_i^*)< \rho-\eps\Big]-o(1)\\
\vspace{5pt} &&\\
&\stackrel{(c)}{=}& 1-o(1).
\end{IEEEeqnarray}
Step (a) uses a union bound. Step (b) is due to Lemma~\ref{lemma:q_prop} and the definition of total variation. Step (c) follows from the hypothesis in \eqref{p2toshow} and the fact that $1/\alpha$ is a constant that does not grow with $n$.

We now show that \eqref{p2toshow} holds for all $j\in[1/\alpha]$. Since our analysis is the same for all sub-blocks, we drop the subscript on $\Bcal_{(j)}$ and simply consider an arbitrary sub-block $\Bcal$ of size $k$.  

We cannot use the standard law of large numbers to show \eqref{p2toshow} because the dependence of $Z^*_i$ on $(M,W^{i-1})$ implies that the random variables $\{\pi(X_i,Y_i,Z_i^*)\}_{i\in \Bcal}$ are not mutually independent. Instead, we condition on $U^n(M)$ and use a martingale argument.

For simplicity, denote $U^n(M)$ by $\ovr{U}^n$. Let $\{S_t\}_{t\in\Bcal}$ be defined by
\begin{equation}
S_t \triangleq \sum_{i\in \Bcal^t} (\pi(X_i,Y_i,Z^*_i)-\rho_i(\ovr{U}^n)),
\end{equation}
where
\begin{equation}
\rho_i(\ovr{U}^n)\triangleq \min_{z(u^n)}\Ebb_{\widehat{Q}}[\pi(X_i,Y_i,z)\: | \: \ovr{U}^n].
\end{equation}

We claim that, conditioned on $\ovr{U}^n$, $S_t$ is a submartingale, i.e., 
\begin{equation}
\Ebb_{\widehat{Q}}[S_t\: | \: S^{t-1},\ovr{U}^n]\geq S_{t-1},\quad \forall t\in \Bcal.
\end{equation}
To verify the claim, first observe that the definition of $S_t$ gives
\begin{IEEEeqnarray}{rCl}
\nonumber\Ebb_{\widehat{Q}}[S_t\mid S^{t-1},\ovr{U}^n]&=&S_{t-1}+\Ebb_{\widehat{Q}}[\pi(X_t,Y_t,Z^*_t)\: | \: S^{t-1},\ovr{U}^n]\\
&&-\>\rho_t(\ovr{U}^n).
\end{IEEEeqnarray}
Moreover, for each $t\in\Bcal$, we have
\begin{IEEEeqnarray}{rCl}
\IEEEeqnarraymulticol{3}{l}{\nonumber
\Ebb_{\widehat{Q}}[\pi(X_t,Y_t,Z^*_t)\: | \: S^{t-1},\ovr{U}^n]
}\\
\vspace{-8pt} \nonumber &\geq& \min_{z(m,w^{t-1},u^n,s^{t-1})} \Ebb_{\widehat{Q}}[\pi(X_t,Y_t,z(M,W^{t-1}))\: | \: S^{t-1},\ovr{U}^n]\\
\vspace{5pt} && \\
&\stackrel{(a)}{=}& \min_{z(u^n,s^{t-1})}\Ebb_{\widehat{Q}}[\pi(X_t,Y_t,z)\: | \: S^{t-1},\ovr{U}^n]\\
&\stackrel{(b)}{=}& \min_{z(u^n)}\Ebb_{\widehat{Q}}[\pi(X_t,Y_t,z)\: | \: \ovr{U}^n]\\
&=& \rho_t(\ovr{U}^n).
\end{IEEEeqnarray}
Step (a) follows by invoking Lemma~\ref{suffstat} after noting that under $\widehat{Q}$ we have the markov chain
\begin{equation}
\label{p2markov}
(X_t,Y_t)-(\ovr{U}^n,S^{t-1})-(M,W^{t-1}).
\end{equation}
Step (b) follows from the markov chain
\begin{equation}
\label{p2markov2}
(X_t,Y_t)-\ovr{U}^n-S^{t-1}.
\end{equation}

Thus, conditioned on $\ovr{U}^n$, we see that $S_t$ is a submartingale. By Doob's decomposition theorem, we can write $S_t=M_t+A_t$, where $M_t$ is a martingale (conditioned on $\ovr{U}^n$) and $A_t$ is an increasing process with $A_1=0$. Therefore, conditioning on $\ovr{U}^n$, we have
\begin{IEEEeqnarray}{rCl}
\IEEEeqnarraymulticol{3}{l}{
\nonumber\Pbb_{\widehat{Q}}\Big[\frac1k \sum_{i\in \Bcal} \pi(X_i,Y_i,Z_i^*)< \frac1k \sum_{i\in \Bcal} \rho_i(\ovr{U}^n)-\eps \:\Big | \: \ovr{U}^n \Big] 
}\\
\quad&=& \Pbb_{\widehat{Q}}[S_k < -k \eps \: | \: \ovr{U}^n]\\
&\leq& \Pbb_{\widehat{Q}}[M_k < -k\eps \: | \: \ovr{U}^n]\\
&=& \Pbb_{\widehat{Q}}\big[M_k-\Ebb_{\widehat{Q}}[M_k] < -k\eps -\Ebb_{\widehat{Q}}[M_k] \: \big | \: \ovr{U}^n \big ]\\
&\stackrel{(a)}{\leq}& \label{causal:variancebound} \frac{\mbox{Var}_{\widehat{Q}}(M_k\: | \: \ovr{U}^n)}{(k\eps +\Ebb_{\widehat{Q}}[S_1])^2},
\end{IEEEeqnarray}
 where (a) follows from Chebyshev's inequality. Now we recursively bound the variance of $M_k$ (conditioned on $\ovr{U}^n)$ by writing
 \begin{IEEEeqnarray}{rCl}
 \IEEEeqnarraymulticol{3}{l}{\nonumber
 \mbox{Var}(M_k \: | \: \ovr{U}^n)
 }\\
 \quad &\stackrel{(a)}{=}& \mbox{Var}(\Ebb[M_k \: | \: M^{k-1}, \ovr{U}^n])\\
 &&+\>\Ebb[\mbox{Var}(M_k \: | \: M^{k-1}, \ovr{U}^n)]\\
 &\leq& \mbox{Var}(\Ebb[M_k \: | \: M^{k-1},\ovr{U}^n])+O(1)\\
 &=& \mbox{Var}(M_{k-1} \: | \: \ovr{U}^n)+O(1).
 \end{IEEEeqnarray}
 Step (a) uses the law of total variance. The recursion implies $\mbox{Var}_{\widehat{Q}}(M_k \: | \: \ovr{U}^n)\in O(k)$, which, together with \eqref{causal:variancebound}, shows
 \begin{equation}
 \lim_{k\to\infty}\Pbb_{\widehat{Q}}\Big[\frac1k \sum_{i\in \Bcal} \pi(X_i,Y_i,Z_i^*)< \frac1k \sum_{i\in \Bcal} \rho_i(\ovr{U}^n)-\eps \: \Big | \: \ovr{U}^n \Big] = 0.
 \end{equation}
 Since this convergence is uniform for all $\ovr{U}^n$, we can take the expectation over random codebooks to get
 \begin{equation}
 \label{causal:almostdone}
 \lim_{k\to\infty}\Ebb_{\Ccal^{(n)}}\,\Pbb_{\widehat{Q}}\Big[\frac1k \sum_{i\in \Bcal} \pi(X_i,Y_i,Z_i^*)< \frac1k \sum_{i\in \Bcal} \rho_i(\ovr{U}^n)-\eps \Big] = 0.
 \end{equation}
Continuing, notice that $\rho_i(\ovr{U}^n)$ can be written as
\begin{IEEEeqnarray}{rCl}
\rho_i(\ovr{U}^n) &=& \min_{z}\Ebb_{\widehat{Q}}[\pi(X_i,Y_i,z)\: | \: \ovr{U}^n]\\
&=& \min_{z}\Ebb_{\widehat{Q}}[\pi(X_i,Y_i,z)\: | \: \ovr{U}_i]\\
&\triangleq & \rho(\ovr{U}_i)
\end{IEEEeqnarray}
because of the markov chain $(X_i,Y_i) - \ovr{U}_i - \ovr{U}^n$ that holds under $\widehat{Q}$. Furthermore, the expected value of $\rho(\ovr{U}_i)$ is
\begin{IEEEeqnarray}{rCl}
\Ebb_{\Ccal^{(n)}}\, \rho(\ovr{U}_i) &=& \Ebb_{\Ccal^{(n)}}\, \Ebb_{\widehat{Q}}[\pi(X_i,Y_i,z)\: | \: \ovr{U}_i]\\
&=& \min_{z(u)} \Ebb_{\Ccal^{(n)}}\, \pi(X_i,Y_i,z(\ovr{U}_i))\\
&\stackrel{(a)}{=}& \min_{z(u)} \Ebb_{\Ccal^{(n)}}\, \pi(X,Y,z(U))\\
&=& \rho
\end{IEEEeqnarray}
where step (a) is due to the fact (readily verified) that $\Ebb_{\Ccal^{(n)}}\, Q_{X_iY_iU_i(M)} = P_{XYU}$. Therefore, because $\ovr{U}^n$ is i.i.d.\ according to $P_U$ (in expectation over the random codebooks), we can invoke the law of large numbers to get
\begin{equation}
\lim_{k\to\infty}\Ebb_{\Ccal^{(n)}}\,\Pbb_{\widehat{Q}}\Big[ \frac1k \sum_{i\in \Bcal} \rho(\ovr{U}_i) > \rho-\eps \Big] = 1.
\end{equation}
This, together with \eqref{causal:almostdone}, yields
\begin{equation}
\lim_{k\to\infty}\Ebb_{\Ccal^{(n)}}\,\Pbb_{\widehat{Q}}\Big[\frac1k \sum_{i\in \Bcal} \pi(X_i,Y_i,Z_i^*)< \rho-2\eps\Big]=0,
\end{equation}
completing the proof of \eqref{p2toshow}. Finally, we invoke Shannon's random coding argument to ensure the existence of a codebook that satisfies the payoff criterion. This concludes the achievability proof of the {\sc whp} payoff criterion.

\subsection{Discussion: Optimal encoding produces artificial noise}
\label{sec:artificialnoise}
The optimal encoding and decoding scheme designed in this section produces an effect that is worth investigating outside of this particular context of rate-distortion theory for secrecy systems.  In particular, consider the most pessimistic disclosure assumption, that $W=(X,Y)$.  In this case, the communication system effectively corrupts the i.i.d.\ information signal $X^n$ with noise by synthesizing a memoryless broadcast channel, with the information source $X^n$ as input, actions at the intended receiver $Y^n$ as one output, and a sequence $U^n$ as the other output observed by the adversary.  The synthesis is accurate in a particular sense relevant to secrecy.  That is, the communication system, which uses public message M and secret key K to facilitate coordination, synthesizes memoryless noise characterized by $P_{YU|X}$ by producing a distribution on $(X^n,Y^n,M)$ such that $P_{X^nY^n|M}$ closely approximates $\prod_{i=1}^n P_{XY|U}(x_i,y_i|u_i(M))$ for a set of statistically typical $u^n(M)$ sequences.  This behavior is revealed by $\widehat{Q}_{MX^nY^n}$ in \eqref{artificialchannel}, which the proof shows to converge to the induced joint distribution of the system in the limit of large $n$.

Let us now consider why this might be an operationally meaningful criterion for synthesizing noise in a secrecy setting.  Consider an adversary who actually does observe a noise-corrupted version of the information signal, such as one of the outputs of a broadcast channel.  As in any probabilistic situation, rational behavior is based on the posterior distribution of the state of the universe given what is known to the individual.  In this situation that means $P_{X^n,Y^n|U^n}$ will dictate the adversary's optimal behavior, regardless of the objective that the adversary is trying to accomplish.  Therefore, a communication system that mimics $P_{X^n,Y^n|U^n}$ will elicit the same behavior by an adversary for the same observed $U^n$ sequence as would occur if the noisy channel was genuine.  Furthermore, if the observed $U^n$ sequence is statistically representative\footnote{Exact characterization of this depends on the specific objectives of the communication system.} of true noisy observations, then the communication system performance in the presence of an adversary will be equivalent to the memoryless broadcast channel that it mimics.

For comparison, consider the work of Winter in \cite{Winter2005}. Although the communication setting and results in \cite{Winter2005} are quite different from ours in that the setting does not have an information source provided by nature, our proof and methods for achievability bear resemblance. There, he considers a distribution on a triple of variables $(X,Y,U)$ and a communication system that generates correlated random variables $X^n$ and $Y^n$ at two different nodes using communication and secret key in the presence of an adversary.  For the sake of comparison, imagine $Y^n$ as a noisy version of $X^n$.  The secrecy criterion in that work is very strong, requiring that the public message reveal no more about the sequences $X^n$ and $Y^n$ than the correlated sequence $U^n$ would, in the sense that $M$ is stochastically degraded from $U^n$ with respect to $(X^n,Y^n)$.  This is stronger than the secrecy criterion we gave in the previous paragraphs, requiring more communication resources as a consequence. However, the noise synthesis achieved by the communication system of this section, even with the weaker secrecy performance implied by \eqref{artificialchannel}, has the same compelling operational significance---an adversary can gain no more advantage from the eavesdropped message than they could by observing the correlated $U^n$ sequence.

\section{Converse Proof}
\label{sec:converse}
It is enough to prove the converse to Theorem~\ref{mainthm} for just the {\sc avg} payoff criterion, since it is the weakest of the criteria. We further weaken the conditions by allowing Node B causal access to Nodes A and C (i.e., we permit decoders of the form $\{P_{Y_i|MKX^{i-1}Z^{i-1}}\}_{i=1}^n$). We will see that this allowance does not increase the payoff. 

Fix a source distribution $P_X$, a payoff function $\pi(x,y,z)$, and causal disclosure channels $P_{W_x|X}$ and $P_{W_y|Y}$. For ease of presentation, denote the pair $(W_x^n,W_y^n)$ by $W^n$. Next, let $J$ be an auxiliary random variable drawn uniformly from $[n]$, independently of $(X^n,Y^n,W^n,M,K)$. Define the following random variables:
\begin{IEEEeqnarray}{rClrCl}
 X &=& X_J \\
 Y &=& Y_J \\
 Z &=& Z_J  \\
 (W_x,W_y) &=& W_J\\
 U &=& (M,W^{J-1},J)\\
 V &=& K.
\end{IEEEeqnarray}
With these choices, it can be verified that
\begin{IEEEeqnarray}{l}
W_x-X-(U,V)-Y-W_y\\
X\sim P_X\\
W_x|X \sim P_{W_x|X}\\
W_y|Y \sim P_{W_y|Y}
\end{IEEEeqnarray}
The following properties of $P_{MKX^nY^nW^n}$ can also be verified:
\begin{IEEEeqnarray}{l}
 \label{convprop1} X^n \perp K\\
 \label{convprop2} X_J - (M,K,X^{J-1},J) - W^{J-1}\\
 \label{convprop3} X_J \perp J.
\end{IEEEeqnarray}
Let $(R,R_0,\Pi)$ be an achievable triple. We first have
\begin{IEEEeqnarray}{rCl}
 nR &\geq& H(M)\\
 &\geq& H(M|K)\\
 &\geq& I(X^n;M|K)\\
 &\stackrel{(a)}{=}& I(X^n;M,K)\\
 &=& \sum_{j=1}^n I(X_j;M,K|X^{j-1})\\
 &=& \sum_{j=1}^n I(X_j;M,K,X^{j-1})\\
 &\stackrel{(b)}{=}& \sum_{j=1}^n I(X_j;M,K,X^{j-1},W^{j-1})\\
 &\geq& \sum_{i=1}^n I(X_j;M,K,W^{j-1})\\
 &\stackrel{(c)}{=}& nI(X_J;M,K,W^{J-1},J)\\
 &=& nI(X;U,V),
\end{IEEEeqnarray}
where (a), (b), and (c) follow from \eqref{convprop1}, \eqref{convprop2}, and \eqref{convprop3}. Next, we have
\begin{IEEEeqnarray}{rCl}
 nR_0 &\geq& H(K)\\
 &\geq& H(K|M)\\
 &\geq& I(W^n;K|M)\\
 &\geq& \sum_{j=1}^n I(W_j;K|M,W^{j-1})\\
 &\stackrel{(a)}{=}& nI(W_J;K|M,W^{J-1},J)\\
 &=& nI(W;V|U),
\end{IEEEeqnarray}
where (a) follows from \eqref{convprop3}. Finally, we have
\begin{IEEEeqnarray}{rCl}
 \Pi &\leq& \min_{z(m,w^{j-1},j)} \Ebb\,\frac1n \sum_{j=1}^n \pi(X_j,Y_j,z(M,W^{j-1},j))\\
 &=& \min_{z(m,w^{j-1},j)} \Ebb\Big[\Ebb[\pi(X_J,Y_J,z(M,W^{J-1},J))|J]\Big]\quad\:\\
 &=& \min_{z(m,w^{j-1},j)} \Ebb\,\pi(X_J,Y_J,z(M,W^{J-1},J))\\
 &=& \min_{z(u)}\Ebb\,\pi(X,Y,z(U)).
 \end{IEEEeqnarray}
It remains to bound the cardinality of $\Ucal$ and $\Vcal$, which is straightforward from the standard support lemma (e.g., \cite{Csiszar2011}). Note that the set of markov distributions forms a compact, connected set. To bound $\Ucal$, it suffices to have $|\Xcal|-1$ elements to preserve $P_X$ and 3 more elements to preserve $H(X|U,V)$, $I(W;V|U)$, and $\min_{z(u)}\Ebb\, \pi(X,Y,z(U))$. To bound $\Vcal$, it suffices to have $|\Xcal||\Ycal||\Ucal|-1$ elements to preserve $P_{XYU}$ and 2 more elements to preserve $H(X|U,V)$ and $H(W|U,V)$.

\section{Other forms of disclosure}
\label{sec:relatedresults}
In this section, we consider several relevant scenarios that are not directly subsumed by Theorem~\ref{mainthm}, but that can be solved by modifying the proof slightly. Throughout, we denote $(W_x^n,W_y^n)$ by $W^n$. Whereas previously we assumed that the eavesdropper has access to causal disclosure $W^{i-1}$, now we consider three other types of disclosure: $W_i$, $W^i$, and $W^n$. It turns out that the regions corresponding to $W^i$ and $W^n$ are the same. 
\begin{thm}
Fix $P_X$, $\pi(x,y,z)$ and disclosure channels $P_{W_x|X}$ and $P_{W_y|Y}$. If $W_i$ is disclosed instead of $W^{i-1}$, then the rate-payoff region for all three payoff criteria is equal to
\begin{equation}
\bigcup_{P_{Y|X}}\left\{
 \begin{IEEEeqnarraybox}[][c]{rCl}
 \IEEEeqnarraymulticol{3}{l}{
 (R,R_0,\Pi): \vspace{2pt}
 }\\
 R &\geq& I(X;Y)\\
 R_0 &\geq& 0\\
 \Pi &\leq& \min_{z(w_x,w_y)} \Ebb\, \pi(X,Y,z(W_x,W_y))
 \end{IEEEeqnarraybox}
\right\}.
\end{equation}
\end{thm}
\begin{IEEEproof}
The proof of achievability is very similar to that of Section~\ref{sec:achievability}. Define the random codebook, encoder, decoder, and $Q_{X^nMKY^nW^n}$ in the same way, but set $U=\emptyset$ and $V=Y$ throughout. Lemma~\ref{lemma:PapproxQ} ensures that the system-induced distribution is approximated by $Q$ since $R > I(X;Y)$. Instead of the property in Lemma~\ref{lemma:q_prop}, the desired property of $Q$ is now
\begin{equation}
Q_{MX_{\Bcal}Y_{\Bcal}W_{\Bcal}} \approx Q_{M} \cdot \Big (  \prod_{i\in\Bcal} P_{XYW}  \Big ).
\end{equation}
The soft covering lemma can be invoked to show that this property holds if the rate of secret key satisfies
\begin{equation}
R_0 > \limsup_{n\to\infty} \frac1n i_Q(X_{\Bcal}Y_{\Bcal}W_{\Bcal};Y^n)=0.
\end{equation}
Thus, under $Q$, the message $M$ is approximately independent of $(X_i,Y_i,W_i)$ and the eavesdropper's best estimate of $(X_i,Y_i)$ only depends on his observation of the disclosure $W_i$. The payoff analysis of Section~\ref{sec:achievability} is straightforward to modify accordingly.

To prove the converse, it is first straightforward to bound $R$ and $R_0$. To bound $\Pi$, define $(W_x,W_y)=W_J$, where $J\sim\mbox{Unif}(n)$, and write
\begin{IEEEeqnarray}{rCl}
 \Pi &\leq& \min_{z(m,w_j,j)} \Ebb\,\frac1n \sum_{j=1}^n \pi(X_j,Y_j,z(M,W_j,j))\\
 &\leq& \min_{z(w_j)} \Ebb\,\frac1n \sum_{j=1}^n \pi(X_j,Y_j,z(W_j))\\
 &=& \min_{z(w)} \Ebb\,\pi(X_J,Y_J,z(W_J))\\
 &=& \min_{z(w_x,w_y)}\Ebb\,\pi(X,Y,z(W_x,W_y)).
 \end{IEEEeqnarray}
\end{IEEEproof}

\begin{thm}
Fix $P_X$, $\pi(x,y,z)$ and disclosure channels $P_{W_x|X}$ and $P_{W_y|Y}$. If $W^n$ or $W^i$ is disclosed instead of $W^{i-1}$, then the rate-payoff region for all three payoff criteria is equal to
\begin{equation}
\bigcup\left\{
 \begin{IEEEeqnarraybox}[][c]{rCl}
 \IEEEeqnarraymulticol{3}{l}{
 (R,R_0,\Pi): \vspace{2pt}
 }\\
 R &\geq& I(X;U,V)\\
 R_0 &\geq& I(W_x,W_y;V|U)\\
 \Pi &\leq& \min_{z(u,w_x,w_y)} \Ebb\, \pi(X,z(U,W_x,W_y))
 \end{IEEEeqnarraybox}
\right\},
\end{equation}
where the union is taken over all markov chains
\begin{equation}
{W_x-X-(U,V)-Y-W_y}.
\end{equation}
\end{thm}
\begin{IEEEproof}
For the proof of achievability, suppose $W^n$ is disclosed. The proof is almost exactly the same as in Section~\ref{sec:achievability}. The code and the rates are identical, as is the definition of the approximating distribution $Q$. Notice that under $\widehat{Q}$ (defined in Lemma~\ref{lemma:q_prop}), the following markov chain holds for all $i\in[n]$: 
\begin{equation}
\label{new_markov}
(X_i,Y_i)-(U_i(M),W_i)-(M,W^n).
\end{equation}
Thus, the eavesdropper's best strategy only depends on $(U_i(M),W_i)$; the rest of the disclosure of $W^n$ is rendered useless. To adjust the analysis of the payoff criteria, simply use markov relations similar to the one in \eqref{new_markov}.

To show the converse proof, suppose that only $W^i$ is disclosed. The proof follows arguments similar to those in Section~\ref{sec:converse}, with exactly the same identification of random variables.
\end{IEEEproof}

\section{Causal disclosure with delay}
\label{sec:delay}
In this section, we consider the effects of assuming that the adversary has delayed causal access to the system behavior. In other words, we replace causal disclosure $W^{i-1}$ with $W^{i-d}$, $d>1$. Surprisingly, this has a major effect on relaxing the amount of secret key required to maintain secrecy. We establish an inner and outer bound on the corresponding rate-payoff region and give an example in which the bounds meet.\footnote{Numerical investigation reveals that the bounds are not tight in general.} Using the bounds, we further show that if lossless communication is required, the minimum rate of secret key needed to ensure a given level of payoff is on the order of $1/d$.
\subsection{Inner and outer bound}
\begin{thm}[Inner bound, causal disclosure with delay $d$]
\label{innerdelaythm}
Fix $P_X$, $\pi(x,y,z)$, and causal disclosure channels $P_{W_x|X}$ and $P_{W_y|Y}$. Let $\Rcal_d$ denote the closure of achievable $(R,R_0,\Pi)$ when the causal disclosure has delay $d\geq 1$. Then

 \begin{equation}
 \label{innerdelaythmregion}
 \Rcal_d \supseteq \bigcup\left\{
 \begin{IEEEeqnarraybox}[][c]{rCl}
 \IEEEeqnarraymulticol{3}{l}{
 (R,R_0,\Pi): \vspace{2pt}
 }\\
  R &\geq& \tfrac1d I(X^d;U,V)\\
 R_0 &\geq& \tfrac1d I(W_x^dW_y^d;V|U)\\
 \Pi &\leq&  \min_{z(u)}\, \Ebb\Big[ \tfrac1d \sum_{j=1}^d\pi(X_j,Y_j,z(U))\Big]
 \end{IEEEeqnarraybox}
\right\},
\end{equation}
where the union is taken over all markov chains
\begin{equation}
{W_x^d-X^d-(U,V)-Y^d-W_y^d}
\end{equation}
in which
\begin{equation}
P_{X^dW_x^d}=\prod_{j=1}^d P_X P_{W_x|X}
\end{equation}
and
\begin{equation}
P_{W_y^d|Y^d}=\prod_{j=1}^d P_{W_y|Y}.
\end{equation}
\end{thm}
\begin{IEEEproof}
For simplicity, we present the proof for $d=2$. Denote $(W_x^n,W_y^n)$ by $W^n$. The idea is to transform the problem into one involving delay $d=1$ so that we can apply Theorem~\ref{mainthm}. To that end, we first treat the source $X^n$ as an i.i.d. sequence $\widetilde{X}^{\frac{n}{2}}$ of super-symbols of length $2$ by defining
\begin{equation}
\widetilde{X}_i = (X_{2i-1},X_{2i}),\,i=1,2,\ldots,n/2.
\end{equation}
Similarly, treat $Y^n$ and $W^n$ as sequences of super-symbols by appropriately defining $\widetilde{Y}^{\frac{n}{2}}$ and $\widetilde{W}^{\frac{n}{2}}$. Under this definition, observe that at steps $i=2,4,\ldots,n$ the adversary has access to $\widetilde{W}^{i-1}=W^{i-2}$. Suppose that at steps $i=1,3,\ldots,n$ we disclose additional information $W_{i-1}$ to the adversary. Now the causal disclosure to the adversary is exactly $\widetilde{W}^{i-1}$ for all $i\in[n]$. Note that supplying extra information to the adversary can only reduce the achievable region.

To complete the transformation, define a payoff function $\widetilde{\pi}:\Xcal^2\times\Ycal^2\times\Zcal^2\rightarrow \Rbb$ by
\begin{equation}
\widetilde{\pi}(x^2,y^2,z^2)=\sum_{j=1}^2\pi(x_j,y_j,z_j).
\end{equation}
If $(\widetilde{R},\widetilde{R}_0,\widetilde{\Pi})$ is an achievable triple for this transformed problem, then $(\widetilde{R}/2,\widetilde{R}_0/2,\widetilde{\Pi}/2)$ is an achievable triple for the delayed causal disclosure problem with $d=2$. By applying Theorem~\ref{mainthm}, we obtain the region in \eqref{innerdelaythmregion} for $d=2$.
\end{IEEEproof}

\begin{thm}[Outer bound, causal disclosure with delay $d$]
\label{outerdelaythm}
Fix $P_X$, $\pi(x,y,z)$, and causal disclosure channels $P_{W_x|X}$ and $P_{W_y|Y}$. Let $\Rcal_d$ denote the closure of achievable $(R,R_0,\Pi)$ when the causal disclosure has delay $d\geq 1$. Then

 \begin{equation}
 \label{outerdelaythmregion}
 \Rcal_d \subseteq \bigcup\left\{
 \begin{IEEEeqnarraybox}[][c]{rCl}
 \IEEEeqnarraymulticol{3}{l}{
 (R,R_0,\Pi): \vspace{2pt}
 }\\
  R &\geq& I(X;U,V)\\
 R_0 &\geq& \tfrac1d I(W_xW_y;V|U)\\
 \Pi &\leq&  \min_{z(u)}\, \Ebb\,\pi(X,Y,z(U))
 \end{IEEEeqnarraybox}
\right\},
\end{equation}
where the union is taken over all markov chains
\begin{equation}
{W_x-X-(U,V)-Y-W_y}.
\end{equation}
\end{thm}
\begin{IEEEproof}
The key to the proof is the following lemma.
\begin{lemma}
\label{outerboundlemma}
For arbitrary random variables $(X^n,Y)$, it holds that
\begin{equation}
d\cdot I(X^n;Y) \geq \sum_{i=1}^n I(X_i;Y|X^{i-d}).
\end{equation}

\end{lemma}
\begin{IEEEproof}[Proof of Lemma~\ref{outerboundlemma}]
\begin{IEEEeqnarray}{rCl}
\IEEEeqnarraymulticol{3}{l}{
\nonumber d\cdot I(X^n;Y) 
}\\
\quad &=& \sum_{j=1}^d I(X^n;Y)\\
&\geq& \sum_{j=1}^d I(X_j^{n-((n-j)\bmod{d})};Y)\\
&\stackrel{(a)}{=}& \sum_{j=1}^d \sum_{i\in[n],i\geq j,i\equiv j\bmod{d}} I(X_{i-d+1}^i;Y|X^{i-d})\\
&=& \sum_{i=1}^nI(X_{i-d+1}^i;Y|X^{i-d})\\
&\geq& \sum_{i=1}^n I(X_i;Y|X^{i-d}).
\end{IEEEeqnarray}
Step (a) uses the chain rule for mutual information on each of the $d$ terms.
\end{IEEEproof}
The converse steps of Section~\ref{sec:converse} can now be modified by defining $U=(M,W^{J-d},J)$.

First, bound $R$ by writing
\begin{IEEEeqnarray}{rCl}
 nR &\geq& H(M)\\
 \nonumber &\vdots& \\
 &\geq& nI(X_J;M,K,W^{J-1},J)\\
 &\geq& nI(X_J;M,K,W^{J-d},J)\\
 &=& nI(X;U,V).
\end{IEEEeqnarray}

Next, bound $R_0$ by writing
\begin{IEEEeqnarray}{rCl}
 d\cdot nR &\geq& d\cdot H(M)\\
 \nonumber &\vdots& \\
 &\geq& d\cdot I(W^n;K|M)\\
 &\stackrel{(a)}{\geq}& \sum_{j=1}^n I(W_j;K|M,W^{j-d})\\
 &=& nI(W_J;K|M,W^{J-d},J)\\
 &=& nI(W;V|U),
\end{IEEEeqnarray}
where (a) uses Lemma~\ref{outerboundlemma}. Finally, $\Pi$ can be bounded in the manner of Section~\ref{sec:converse}.
\end{IEEEproof}
\subsection{Lossless communication}
We now specialize the inner and outer bound to the setting in which lossless communication is required and $X^{i-d}$ is disclosed. In this regime, we are able to show explicitly how delay affects the tradeoff between rate of secret key and payoff.
\begin{thm}
Fix $P_X$ and $\pi(x,z)$. Let $\Rcal_d$ denote the closure of achievable $(R,R_0,\Pi)$ for the case of lossless communication and causal disclosure $X^{i-d}$, $d\geq 1$. Let $R_d(\Pi)$ denote the key-payoff boundary of $\Rcal_d$. First, we have
\begin{equation}
 \label{losslessXinnerdelay}
 \Rcal_d \supseteq \bigcup_{\substack{P_{X^dU}: \\X^d\sim \prod_{j=1}^d P_X}}\left\{
 \begin{IEEEeqnarraybox}[][c]{rCl}
 \IEEEeqnarraymulticol{3}{l}{
 (R,R_0,\Pi): \vspace{2pt}
 }\\
 R &\geq& H(X)\\
 R_0 &\geq& \tfrac 1d H(X^d|U)\\
 \Pi &\leq& \min_{z(u)} \Ebb\Big[\tfrac1d \sum_{j=1}^d \pi(X_j, z(U))\Big]
 \end{IEEEeqnarraybox}
\right\}
\end{equation}
and
\begin{equation}
 \label{losslessXouterdelay}
 \Rcal_d \subseteq \bigcup_{\substack{P_{XU}:\\X\sim P_X}}\left\{
 \begin{IEEEeqnarraybox}[][c]{rCl}
 \IEEEeqnarraymulticol{3}{l}{
 (R,R_0,\Pi): \vspace{2pt}
 }\\
 R &\geq& H(X)\\
 R_0 &\geq& \tfrac1d H(X|U)\\
 \Pi &\leq& \min_{z(u)} \Ebb\, \pi(X, z(U))
 \end{IEEEeqnarraybox}
\right\}.
\end{equation}
Furthermore, for all $\Pi$,
\begin{equation}
\label{delayorder}
R_d(\Pi)=\Theta\Big(\frac1d\Big).
\end{equation}
\end{thm}
\begin{IEEEproof}
To establish the inner bound on $\Rcal_d$, first recall the characterization of $\Rcal_{1}$ given in Corollary~\ref{losslessX}. Using the same arguments as the proof of Theorem~\ref{innerdelaythm}, we can transform the problem with delay $d>1$ into one involving delay $d=1$ and invoke Corollary~\ref{losslessX} on the new problem. Upon noting that $\tfrac1d H(X^d)=H(X)$ when $X^d \sim \prod {P_X}$, this technique gives the achievable region in \eqref{losslessXinnerdelay}.

To establish the outer bound, let $(R,R_0,\Pi)$ be an achievable triple. The bound $R\geq H(X)$ is due to the lossless source coding theorem. To bound $R_0$ and $\Pi$, let $J$ be uniformly distributed on $[n]$ and define $U=(M,X^{J-d},J)$ and $X=X_J$. Then, we have
\begin{IEEEeqnarray}{rCl}
nR_0 &\geq& nH(K) \\
&\geq& nI(X^n;K|M) \\
&=& nH(X^n|M) - nH(X^n|K,M) \\
&\stackrel{(a)}{=}& nH(X^n|M) - n\cdot o(1) \\
&\stackrel{(b)}{\geq}& n\cdot \tfrac1d \sum_{j=1}^n H(X_j | X^{j-d}, M) - n\cdot o(1) \\
&=& n\cdot \tfrac1d H(X_J | X^{J-1},M,J) - n\cdot o(1) \\
&=& n\cdot \tfrac1d H(X|U) - n\cdot o(1).
\end{IEEEeqnarray}
Step (a) uses Fano's inequality, and step (b) follows from Lemma~\ref{outerboundlemma} (by setting $Y=X^n$ and conditioning on $M$). It is straightforward to bound $\Pi$ in the manner of Section~\ref{sec:converse}.

From the outer bound in \eqref{losslessXouterdelay}, we see that
$R_d(\Pi) \geq \tfrac1d R_1(\Pi)$. It remains to show that $R_d(\Pi)\leq c\cdot\frac1d$ for some constant $c$; we do this via \eqref{losslessXinnerdelay}. First, let $X^d\sim\prod_{j=1}^d P_X$. Let $K\sim\mbox{Unif}(\Xcal)$ be independent of $X^d$ and define
\begin{equation}
\label{auxUdefn}
U\triangleq(X_1\oplus K,X_2\oplus K,\ldots,X_d\oplus K),
\end{equation}
where $\oplus$ indicates addition modulo $\Xcal$. With this choice of $U$, we have
\begin{equation}
\label{delayexampleprop1}
H(X_i|X_j,U)=0,\,\forall i,j\in[d]
\end{equation}
and
\begin{equation}
\label{delayexampleprop2}
X_j \perp U, \,\forall j\in[d].
\end{equation}
Therefore, we can write
\begin{IEEEeqnarray}{rCl}
\tfrac1d H(X^d|U) &=& \tfrac1d \sum_{i=1}^d H(X_j|X^{j-1},U)\\
&\stackrel{(a)}{=}& \tfrac1d H(X_1|U)\\
&\stackrel{(b)}{=}& \tfrac1d H(X),
\end{IEEEeqnarray}
where (a) and (b) follow from \eqref{delayexampleprop1} and \eqref{delayexampleprop2}, respectively. Moreover, we have
\begin{IEEEeqnarray}{rCl}
\IEEEeqnarraymulticol{3}{l}{
\min_{z(u)} \Ebb\Big[\tfrac1d \sum_{j=1}^d \pi(X_j, z(U))\Big]
}\\
\quad &=& \tfrac1d \sum_{j=1}^d \min_{z(u)} \Ebb\, \pi(X_j, z(U))\\
&\stackrel{(a)}{=}& \min_{z} \Ebb\, \pi(X, z)\\
&\triangleq& \pi_{\max},
\end{IEEEeqnarray}
where (a) follows from Lemma~\ref{suffstat} and \eqref{delayexampleprop2}. 

By selecting $U$ according to \eqref{auxUdefn}, we have shown that the inner bound in \eqref{losslessXinnerdelay} contains the point $(R_0,\Pi)=(\frac1d H(X), \pi_{\max})$; therefore, $(\frac1d H(X), \pi_{\max})\in\Rcal_d$. Since $\pi_{\max}$ is the maximum possible payoff, this implies $R_d(\Pi)\leq \frac1d H(X)$, completing the proof of \eqref{delayorder}.
\end{IEEEproof}
\subsection{Example in which the bounds meet}
In the preceding proof, we demonstrated that the point $(R_0,\Pi)=(\frac1d H(X),\pi_{\max})$ is in the region \eqref{losslessXinnerdelay} and is therefore achievable. If we choose the source distribution to be $P_X\sim\mbox{Bern}(1/2)$, then from Theorem~\ref{losslessexample} (which gives us $R_1(\Pi)$) and the convexity of the rate-payoff region, it is clear that $R_d(\Pi) \leq \frac1d R_1(\Pi)$. Conversely, the outer bound in \eqref{losslessXouterdelay} directly gives $R_d \geq \frac1d R_1(\Pi)$.

\section{Conclusion}
This work has established a theory of secure source coding which characterizes the optimal use of communication and secret key to allow good reconstruction of the source by the intended receiver (who has access to the key) and force a poor reconstruction on any eavesdropper (without the secret key).  The central contribution, presented in Theorem~\ref{mainthm}, gives a general information theoretic characterization of the achievable performance.  The expression in the theorem makes use of two auxiliary variables which can be interpreted as information that is kept secure and information that is released publicly.  In the case of lossless compression in Corollary~\ref{losslessX}, the optimal communication system can explicitly follow these implied steps, constructing two separate messages and focusing all of the security resource (i.e., the key) on only one.

An important component of the main result is the causal disclosure assumption depicted in Fig.~\ref{fig:model}, which was absent from YamamotoÕs formulation of the problem in \cite{Yamamoto1988} and \cite{Yamamoto1997}.  The causal disclosure empowers the eavesdropper with additional information and forces the communication system to resort to a more robust design for secure encoding, which results in an innovative encoding and decoding scheme that sterilizes the causal disclosure.

The theorems in this work allow for an arbitrary but known disclosure channel to the eavesdropper.  However, one could always take the most pessimistic approach and assume that the source $X$ and the reconstruction $Y$ are both fully disclosed (causally) to the eavesdropper.  This leads to the strongest definition of secrecy in our model, and the optimal communication system for this setting has a simple and natural interpretation as producing synthetic noise, discussed in Section~\ref{sec:artificialnoise}.

This work also identifies the rate-distortion tradeoffs without the causal disclosure assumption.  The case of no disclosure (as in Yamamoto's model) is a special case of the main result and is addressed in Section~\ref{sec:onebit}, along with a discussion of its fragility.  Non-causal disclosure is the topic of Section~\ref{sec:relatedresults}, which turns out to only be as empowering to the eavesdropper as causal disclosure.

The causal disclosure framework boasts some important unique properties aside from its operational interpretation as real-time reconstruction by the eavesdropper.  In Section~\ref{sec:equivocation} we show that the traditional approach of measuring secrecy by normalized equivocation (rather than distortion) is in fact a special case of this framework by applying a particular log-loss distortion function.  This connection only exists because of the causal disclosure assumption.  Another property that arises is the need for a stochastic decoder, which suggests a duality with Wyner's wiretap channel \cite{Wyner1975wiretap} where a stochastic encoder is needed.  Furthermore, this framework induces a rich tradeoff between the rate of secret key used and the distortion the system imposes upon an eavesdropper, while such a tradeoff does not occur in the absence of causal disclosure.  These features suggest that causal disclosure is an appropriate base assumption for understanding rate-distortion theory for secrecy systems.

\appendices
\section{Proof of Theorem~\ref{losslessexample}}
\label{losslessexampleappendix}

\subsection{Supporting lemma}
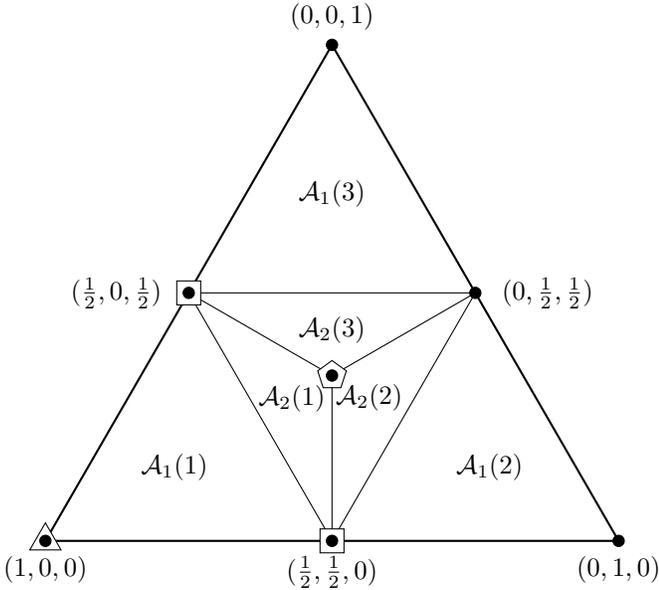
\begin{figure}
\centering
\begin{tikzpicture}[scale=4.4,dot/.style={draw=black,fill=black,circle,minimum size=1.5mm,inner sep=0pt}]

\node[coordinate] (ctr) at (0,0) {};

\node[coordinate] (A) at (210:1cm) {};
\node[coordinate] (B) at (-30:1cm) {};
\node[coordinate] (C) at (90:1cm) {};

\node[coordinate] (AB) at ($(A)!0.5!(B)$) {};
\node[coordinate] (BC) at ($(B)!0.5!(C)$) {};
\node[coordinate] (AC) at ($(A)!0.5!(C)$) {};

\node[rectangle,inner sep = 0pt] [above=2mm of C] {$(0,0,1)$};
\node[rectangle,inner sep =0pt] [left =3.5mm of AC] {$(\tfrac12,0,\tfrac12)$};
\node[rectangle,inner sep =0pt] [right =3.5mm of BC] {$(0,\tfrac12,\tfrac12)$};
\node[rectangle,inner sep =0pt] [below =2mm of A] {$(1,0,0)$};
\node[rectangle,inner sep =0pt] [below =2mm of AB] {$(\tfrac12,\tfrac12,0)$};
\node[rectangle,inner sep =0pt] [below =2mm of B] {$(0,1,0)$};

\node[rectangle,inner sep =0pt] at (210:5.5mm) {$\mathcal{A}_1(1)$};
\node[rectangle,inner sep =0pt] at (210:1.4mm) {$\mathcal{A}_2(1)$};

\node[rectangle,inner sep =0pt] at (-30:5.5mm) {$\mathcal{A}_1(2)$};
\node[rectangle,inner sep =0pt] at (-30:1.3mm) {$\mathcal{A}_2(2)$};

\node[rectangle,inner sep =0pt] at (90:5.5mm) {$\mathcal{A}_1(3)$};
\node[rectangle,inner sep =0pt] at (90:1.4mm) {$\mathcal{A}_2(3)$};

\draw[thick] (A) -- (B) -- (C) -- (A);

\draw (AB) -- (BC) -- (AC) -- (AB);

\draw (ctr) -- (AB);
\draw (ctr) -- (BC);
\draw (ctr) -- (AC);

\node[draw, shape border rotate=0, regular polygon, regular polygon sides=3, inner sep =0pt,minimum height=4.8mm,fill=white] at (A) {};
\node[draw, shape border rotate=0, regular polygon, regular polygon sides=4, inner sep =0pt,minimum height=4.5mm,fill=white] at (AB) {};
\node[draw, shape border rotate=0, regular polygon, regular polygon sides=4, inner sep =0pt,minimum height=4.5mm,fill=white] at (AC) {};
\node[draw, shape border rotate=0, regular polygon, regular polygon sides=5, inner sep =0pt,minimum height=4mm,fill=white] at (0,0) {};

\node[dot] at (ctr) {};

\node[dot] at (A) {};
\node[dot] at (B) {};
\node[dot] at (C) {};

\node[dot] at (AB) {};
\node[dot] at (BC) {};
\node[dot] at (AC) {};

\end{tikzpicture}
\caption[]{\small The probability simplex $\Delta_{\mathcal{X}}$ for $\mathcal{X}=\{1,2,3\}$. The centroid is the distribution $(\tfrac13,\tfrac13,\tfrac13)$. Note that $\mathcal{F}_1(1)=\{  \tikz{\node[draw, shape border rotate=0, regular polygon, regular polygon sides=3, inner sep =0pt,minimum height=3mm,fill=white] {};}  \}$, $\mathcal{F}_2(1)=\{  \tikz{\node[draw, shape border rotate=0, regular polygon, regular polygon sides=4, inner sep =0pt,minimum height=3mm,fill=white] {};} \}$, and $\mathcal{F}_3(1)=\{  \tikz{\node[draw, shape border rotate=0, regular polygon, regular polygon sides=5, inner sep =0pt,minimum height=3mm,fill=white] {};}  \}$.}
\label{fig:simplex}
\end{figure}

For each $x\in\Xcal$, define $\Fcal_n(x)\subseteq\Delta_{\Xcal}$ by
\begin{IEEEeqnarray}{l}
\nonumber \Fcal_n(x)\triangleq \Big\{p\in\Delta_{\Xcal}: p=\mbox{Unif}(\Acal) \mbox{ for some }\Acal\subseteq \Xcal, \\
\qquad \qquad \quad |\Acal|=n,\mbox{ and } p(x)=\max_{x'} p(x') \Big\},
\end{IEEEeqnarray}
and define $\Acal_n(x)\subseteq\Delta_{\Xcal}$ by
\begin{IEEEeqnarray}{l}
\nonumber \Acal_n(x)\triangleq \Big\{p\in\Delta_{\Xcal}: p(x)=\max_{x'} p(x')\\
\qquad\qquad\quad \mbox { and } p(x)\in\left[\tfrac{1}{n+1},\tfrac1n\right] \Big\}.
\end{IEEEeqnarray}
In words, $\Fcal_n(x)$ is the set of probability mass functions on $\Xcal$ that are uniformly distributed on a subset of size $n$ and whose largest mass occurs at $x$. Fig.~\ref{fig:simplex} illustrates the definitions of $\Fcal_n(x)$ and $\Acal_n(x)$ when $\Xcal=\{1,2,3\}$.

The key to the proof of Theorem~\ref{losslessexample} is the following technical lemma.

\begin{lemma}
\label{supportlosslessexample}
For a random variable $X$ with distribution $P_{X}$, let $\ovr{x}$ and $N$ be such that $P_{X}\in \Acal_N(\ovr{x})$.
\begin{enumerate}
\item There exists a random variable $V$, correlated with $X$, such that for all $v\in\Vcal$,
\begin{equation}
P_{X|V=v}\in \Fcal_N(\ovr{x})\cup \Fcal_{N+1}(\ovr{x}).
\end{equation}
In other words, $P_X$ can be written as a convex combination of distributions in $\Fcal_N(\ovr{x})\cup \Fcal_{N+1}(\ovr{x})$.
\item Let $n\in[N]$. There exists a random variable $V$ such that for all $v\in\Vcal$,
\begin{equation}
P_{X|V=v}\in \bigcup_{x\in\Xcal}\Fcal_n(x).
\end{equation}

In other words, for any $n\in[N]$, $P_X$ can be written as a convex combination of distributions in $\cup_x \Fcal_n(x)$.
\end{enumerate}
\end{lemma}

\begin{IEEEproof}
Fix $\xbar\in\Xcal$ and $n\in \Nbb$, and define
\begin{equation}
\Fcal \triangleq \Fcal_n(\xbar)\cup \Fcal_{n+1}(\xbar).
\end{equation}
 First, one can verify that $\Acal_n(\xbar)$ is a convex set. 
 Furthermore, it is well-known that every compact convex set is the convex hull of its extreme points. Thus, to prove part 1, it is enough to show that the set of extreme points of $\Acal_n(\xbar)$ is equal to $\Fcal$. Then any $p\in\Acal_n(\xbar)$ can be written as a convex combination of the elements of $\Fcal$. 
 
 The set of extreme points of a convex set $\Ccal$ is defined by
 \begin{IEEEeqnarray}{l}
 \nonumber \mbox{extr}(\Ccal) \triangleq \{p\in\Ccal: \mbox{if } p=\theta q + (1-\theta) r,\: q,r\in\Ccal,\\
 \qquad\qquad\qquad\qquad \theta\in(0,1) \mbox{ then } p=q=r\}.
 \end{IEEEeqnarray}
 
 We first show that $\Fcal\subseteq\mbox{extr}(\Acal_n(\xbar))$. Let $p\in\Fcal$, and let $q,r\in\Acal_n(\xbar)$, $\theta\in(0,1)$ be such that $q\neq p$, $r\neq p$, and
 \begin{equation}
 p= \theta q + (1-\theta) r
 \end{equation}
If $p\in \Fcal_n(\xbar)$, then $p=q=r$ is clear because $q(x)\in[0,\frac1n]$ and $r(x)\in[0,\frac1n]$ for all $x\in\Xcal$. On the other hand, suppose $p\in \Fcal_{n+1}(\xbar)$. Because $q,r\in\Acal_n(\xbar)$ and $p(\xbar)=\frac{1}{n+1}$, we have $q(\xbar)=r(\xbar)=\frac{1}{n+1}$. Thus, $q(x)\in[0,\frac{1}{n+1}]$ and $r(x)\in[0,\frac{1}{n+1}]$ for all $x\in\Xcal$, and again $p=q=r$.

To show $\mbox{extr}(\Acal_n(\xbar)) \subseteq \Fcal$, we proceed by way of contradiction and suppose that $p\in \mbox{extr}(\Acal_n(\xbar))$ and $p\notin\Fcal$. From $p\notin\Fcal$, it holds that $p(x')\in (0,\frac{1}{n+1})\cup(\frac{1}{n+1},\frac1n)$ for some $x'\in\Xcal$. There are now three separate cases to consider depending on whether $p(\xbar)=\frac{1}{n+1}$, $p(\xbar)\in(\frac{1}{n+1},\frac1n)$, or $p(\xbar)=\frac1n$. For ease of exposition, we only consider $p(\xbar)=\frac{1}{n+1}$; the other two cases use a similar argument. Since $p(x')\leq p(\xbar)$, we have $p(x')\in (0,\frac{1}{n+1})$. It follows that there must exist $x''\neq x'$ such that $p(x'')\in(0,\frac{1}{n+1})$; otherwise, we would have
\begin{equation}
\sum_{x\in\Xcal}p(x)=\tfrac{n}{n+1}+p(x') < 1
\end{equation}
Now we can write $p=\frac12 q+\frac12 r$, where
\begin{equation}
q(x)=
\begin{cases}
p(x), & x\neq x',x\neq x''\\
p(x)+\eps, & x=x'\\
p(x)-\eps, & x=x''
\end{cases}
\end{equation}
\begin{equation}
r(x)=
\begin{cases}
p(x), & x\neq x',x\neq x''\\
p(x)-\eps, & x=x'\\
p(x)+\eps, & x=x''
\end{cases}
\end{equation}
and 
\begin{equation}
\eps=\tfrac12 \min\Big\{p(x'),p(x''),\tfrac{1}{n+1}-p(x'),\tfrac{1}{n+1}-p(x'')\Big\}.
\end{equation}
Thus, $p\notin\mbox{extr}(\Acal_n(\xbar))$, giving the contradiction. We have shown $\Fcal=\mbox{extr}(\Acal_n(\xbar))$ and part 1 of the lemma.

To prove part 2 of the lemma, first define
\begin{equation}
\Bcal_n\triangleq\bigcup_{x\in\Xcal}\Fcal_n(x).
\end{equation}
 For any $n$, it holds that
\begin{equation}
\label{part2fact}
\Bcal_{n+1}\subseteq \mbox{conv}(\Bcal_n).
\end{equation}
This follows from writing $p\in \Bcal_{n+1}$ as
\begin{equation}
p=\sum_{q\in \Bcal_n:\mbox{supp}(q)\subseteq \mbox{supp}(p)}\tfrac{1}{n+1}\,q.
\end{equation}
One can establish part 2 by using part 1 and \eqref{part2fact}.
\end{IEEEproof}
\subsection{Proof of Theorem~\ref{losslessexample}}
With Lemma~\ref{supportlosslessexample} in hand, we are equipped to prove Theorem~\ref{losslessexample}. Fix $R_0$ and let $U^*$ be the maximizer of $\Pi(R_0)$. When the payoff function is $\pi(x,z)=\mathbf{1}\{x \neq z\}$, we can rewrite $\Pi(R_0)$ as
\begin{IEEEeqnarray}{rCl}
\Pi(R_0) &=& \min_{z(u)}\Ebb\,\pi(X,z(U^*)) \\
  \vspace{-10pt} \nonumber&=& \min_{z(u)}\sum_uP_{U^*}(u)\sum_xP_{X|U^*}(x|u)\mathbf{1}\{x\neq z(u)\}\\
  \vspace{5pt}&&\\
 &=& \sum_uP_{U^*}(u)\min_z\sum_xP_{X|U^*}(x|u)\mathbf{1}\{x\neq z\}\\
 &=& \sum_uP_{U^*}(u)\min_z (1-P_{X|U^*}(z|u))\\
 &=& \label{hammingrewrite} \sum_uP_{U^*}(u)(1-\max_xP_{X|U^*}(x|u)).
\end{IEEEeqnarray}
We now show that the set $\{P_{X|U^*=u}\}_u$ in \eqref{tradeoffrewrite} can be restricted the finite set $\Pcal_{\sf {unif}}$, where
\begin{equation}
\Pcal_{\sf unif} \triangleq \{ p\in \Delta_\Xcal: p=\mbox{Unif}(\Acal) \mbox{ for some }\Acal \subseteq \Xcal \}.
\end{equation}
By applying part 2 of Lemma~\ref{supportlosslessexample} to each distribution in $\{P_{X|U^*=u}\}_u$, we have that there exists a random variable $V$ such that
\begin{IEEEeqnarray}{l}
\forall u,v,\: P_{X|U^*=u,V=v}\in\Pcal_{\sf unif}\\
\nonumber \forall u,v,v',\: \arg\max_x P_{X|U^*V}(x|u,v)\\
\label{maxesmatch} \qquad\qquad=\arg\max_x P_{X|U^*V}(x|u,v').
\end{IEEEeqnarray}
We now write
\begin{IEEEeqnarray}{rCl}
\IEEEeqnarraymulticol{3}{l}{
 \nonumber \Pi(R_0)
 }\\
  &\stackrel{(a)}{=}& \sum_uP_{U^*}(u)(1-\max_xP_{X|U^*}(x|u))\\
  \vspace{-10pt} \nonumber &=& \sum_uP_{U^*}(u)(1-\max_x\sum_v P_{X|U^*V}(x|u,v)P_{V|U^*}(v|u))\\
  \vspace{5pt} &&\\
  &\stackrel{(b)}{=}& \sum_{u,v}P_{U^*V}(u,v)(1-\max_xP_{X|U^*V}(x|u,v))\\
  &=& \min_{z(u,v)} \Ebb \, \pi(X,z(U^*,V)),
\end{IEEEeqnarray}
where (a) is due to \eqref{hammingrewrite} and (b) follows from \eqref{maxesmatch}. By noting that $R_0\geq H(X|U^*)\geq H(X|U^*,V)$ and letting $U=(U^*,V)$, we have
\begin{equation}
\Pi(R_0) \leq \max_{\substack{U:\: P_{X|U=u}\in\Pcal_{\sf unif} \\ R_0\geq H(X|U)}} \min_{z(u)} \Ebb\,\pi(X,z(U)).
\end{equation}
This shows that we can restrict attention to $\Pcal_{\sf unif}$ without hurting the payoff. Now, observe that $p\in\Pcal_{\sf unif}$ satisfies
\begin{equation}
\big(H(p),1-\max_x p(x)\big)=\big(\log n, \tfrac{n-1}{n}\big)
\end{equation}
for some $n\in\Nbb$. Referring to \eqref{hammingrewrite} and noting that $H(X|U)=\sum_uP_U(u)H(X|U=u)$, we see that $\Pi(R_0)$ cannot lie outside of the convex hull of the pairs $(\log n, \frac{n-1}{n}),n\in\Nbb$. That is,
\begin{equation}
\Pi(R_0)\leq \phi(R_0).
\end{equation}
To see $\Pi(R_0)\leq \pi_{\max}$, simply write
\begin{IEEEeqnarray}{rCl}
\Pi(R_0) &=& \sum_uP_{U^*}(u)(1-\max_xP_{X|U^*}(x|u))\\
&\leq& 1-\max_{x} \sum_{u} P_{U^*}(u)P_{X|U^*}(x|u)\\
&=& \pi_{\max}.
\end{IEEEeqnarray}

It remains to show that $\min\{\phi(R_0),\pi_{\max}\}$ can be achieved through the proper choice of $U$. To that end, let $\ovr{x}$ and $N$ be such that $P_X\in \Acal_N(\ovr{x})$. By the convexity of $\Rcal$, we will be done once we show that we can achieve not only the points $(\log n, \frac{n-1}{n}),n\in[N]$, but also the intersection of $\phi$ with $\pi_{\max}$. To achieve the point $(\log n, \frac{n-1}{n})$, invoke part 2 of Lemma~\ref{supportlosslessexample} produce $U$. Denote the corresponding rate-payoff pair by $(R_0',\Pi')_n$. Since the $\{P_{X|U=u}\}_u$ all satisfy
\begin{equation}
\big(H(X|U=u),1-\max_x P_{X|U=u}(x|u)\big)=\big(\log n, \tfrac{n-1}{n}\big)
\end{equation}
so must $(R_0',\Pi')_n$ as well. To achieve the intersection of $\phi$ with $\pi_{\max}$, first invoke part 1 of Lemma~\ref{supportlosslessexample} to produce $U$. Denote the corresponding rate-payoff pair by $(R_0'',\Pi'')$. The $\{P_{X|U=u}\}_u$ correspond to either $(\log n, \frac{n-1}{n})$ or $(\log (n+1), \frac{n}{n+1})$. Thus, $(R_0'',\Pi'')$ lies on $f$ because it is a convex combination of those two points. We also have that $(R_0'',\Pi'')$ satisfies $\Pi''=\pi_{\max}$ because 
\begin{equation}
\arg\max_x P_{X|U=u}(x|u)=\ovr{x},\:\forall u\in\Ucal.
\end{equation}
This completes the proof of Theorem~\ref{losslessexample}.

\section{Proof of Lemma~\ref{lemma:q_prop}}
\label{cloudmixingappendix}
Let $R_0 > I(W;V|U)$. Define the typical set
\begin{equation}
\Tcal_\eps^n\triangleq\{u^n:|T_{u^n}(u)-P_U(u)|<\eps P_U(u),\forall u\in\Ucal\}.
\end{equation}
where $T_{u^n}$ denotes the type of $u^n$.

First, write
\begin{IEEEeqnarray}{rCl}
\IEEEeqnarraymulticol{3}{l}{\nonumber
\big\lVert Q_{MW^nX_{\Bcal}Y_{\Bcal}} - \widehat{Q}_{MW^nX_{\Bcal}Y_{\Bcal}}\big\rVert
}\\
\nonumber &=& \hspace{-5pt} \sum_{m:U^n(m)\in\Tcal^n_{\eps}} \hspace{-10pt} Q_{M}(m) \big\lVert Q_{W^nX_{\Bcal}Y_{\Bcal}|M=m} - \widehat{Q}_{W^nX_{\Bcal}Y_{\Bcal}|M=m}\big\rVert\\
\label{twotermsappendix} && + \hspace{-5pt} \sum_{m:U^n(m)\notin\Tcal^n_{\eps}} \vspace{-10pt} \nonumber \hspace{-10pt} Q_{M}(m) \big\lVert Q_{W^nX_{\Bcal}Y_{\Bcal}|M=m} - \widehat{Q}_{W^nX_{\Bcal}Y_{\Bcal}|M=m}\big\rVert.\\
~
\end{IEEEeqnarray}
The expected value of the second term in \eqref{twotermsappendix} can be bounded easily.  For sufficiently large $n$, we have
\begin{IEEEeqnarray}{rCl}
\IEEEeqnarraymulticol{3}{l}{\nonumber
\Ebb \hspace{-10pt} \sum_{m:U^n(m)\notin\Tcal^n_{\eps}} Q_{M}(m) \left\lVert P_{X^nY^k|M=m} - \widehat{Q}_{X^nY^k|M=m} \right\rVert
}\\
\quad&\leq& \Ebb \sum_{m:U^n(m)\notin\Tcal^n_{\eps}} Q_{M}(m) \\
&=& \Pbb[U^n(M)\notin \Tcal^n_{\eps}]\\
&=& \Pbb[U^n(1)\notin \Tcal^n_{\eps}]\\
&\stackrel{(a)}{\leq}& \eps,
\end{IEEEeqnarray}
where (a) is due to the law of large numbers. 

The expected value of the first term in \eqref{twotermsappendix} can first be rewritten by moving the expectation with respect to the subcodebook $C_V^{(n)}(m)$ inside the sum.
\begin{IEEEeqnarray}{rCl}
\IEEEeqnarraymulticol{3}{l}{\nonumber
\Ebb \hspace{-5pt} \sum_{m:U^n(m)\in\Tcal^n_{\eps}} \hspace{-10pt} Q_{M}(m) \left\lVert Q_{W^nX_{\Bcal}Y_{\Bcal}|M=m} - \widehat{Q}_{W^nX_{\Bcal}Y_{\Bcal}|M=m} \right\rVert
}\\
 \nonumber &=& \Ebb _{\Ccal_U^{(n)}} \hspace{-10pt} \sum_{m:U^n(m)\in\Tcal^n_{\eps}} \hspace{-10pt} Q_{M}(m) \,\Ebb _{\Ccal_{V}^{(n)}(m)} \Big\lVert Q_{W^nX_{\Bcal}Y_{\Bcal}|M=m} \\
\label{toshowcloudmixing} && \hspace{4.2cm} -\> \widehat{Q}_{W^nX_{\Bcal}Y_{\Bcal}|M=m} \Big\rVert.
\end{IEEEeqnarray}
It remains to show that the inner expectation vanishes for each $m$. \footnote{Due to the symmetry of codebook construction, the behavior of the inner expectation is uniform for all $m$. Thus, the rate of convergence does not play a role in claiming that \eqref{toshowcloudmixing} vanishes.} 

To do this, first observe that $Q_{W^nX_{\Bcal}Y_{\Bcal}|M=m}$ is the output of the memoryless (but nonstationary) channel $\Phi \triangleq Q_{W^nX_{\Bcal}Y_{\Bcal}|K,M=m}$ acting on a codebook of size $2^{nR_0}$ that is generated i.i.d.\ according to $\Psi \triangleq \prod_i P_{V|U=u_i(m)}$. Furthermore, it can verified that $\widehat{Q}_{W^nX_{\Bcal}Y_{\Bcal}|M=m}$ is the output distribution of the channel $\Phi$ when the input distribution is $\Psi$. Thus, we can invoke the soft covering lemma (Lemma~\ref{generalcloudmixing}) as long as $R_0$ exceeds the sup-information rate of the process that results from $\Phi$ acting on $\Psi$. To be explicit, that process is given by
\begin{IEEEeqnarray}{l}
\nonumber \Gamma(v^n,w^n,x_{\Bcal},y_{\Bcal})\\
\vspace{-10pt} \nonumber \triangleq \prod_{i=1}^n P_{VW|U}(w_i,v_i|u_i(m)) \prod_{i\in\Bcal} P_{VXY|U}(x_i,y_i,v_i|u_i(m)). \\
~
\end{IEEEeqnarray}
Since $\Gamma$ is a memoryless process and the second moments of $\{i_{\Gamma}(W_i,X_i,Y_i;V_i)\}$ are uniformly bounded, the law of large numbers gives
\begin{equation}
\limsup_{n\to\infty} \frac1n i_{\Gamma}(W^n,X_{\Bcal},Y_{\Bcal};V^n) \leq \Ebb\, \frac1n i_{\Gamma}(W^n,X_{\Bcal},Y_{\Bcal};V^n).
\end{equation}
Furthermore, we can upper bound the expected information density by writing
\begin{IEEEeqnarray}{rCl}
\IEEEeqnarraymulticol{3}{l}{\nonumber
\Ebb\,\frac1n i_{\Gamma}(W^n,X_{\Bcal},Y_{\Bcal};V^n)
}\\
\quad &=& \Ebb\,\frac1n i_{\Gamma}(W^n;V^n) + \Ebb\,\frac1n i_{\Gamma}(X_{\Bcal},Y_{\Bcal};V^n|W^n)\\
&=&  \Ebb\,\frac1n i_{\Gamma}(W^n;V^n) + \frac1n I_{\Gamma}(X_{\Bcal},Y_{\Bcal};V^n|W^n)\\
&\leq& \Ebb\,\frac1n i_{\Gamma}(W^n;V^n) + \alpha \log |\Xcal||\Ycal|\\
\nonumber &=&\Ebb\,\frac1n \sum_{i=1}^n i_{P_{WV|U=u_i(m)}}(W;V|U=u_i(m))\\
&& +\>\alpha \log |\Xcal||\Ycal|\\
&=& \frac1n \sum_{i=1}^n I(W;V|U=u_i(m)) + \alpha \log |\Xcal||\Ycal|\\
&=& \sum_{u\in\Ucal} T_{u^n(m)} I(W;V|U=u)+\alpha \log |\Xcal||\Ycal|\\
\vspace{-10pt} \nonumber &\stackrel{(a)}{\leq}& \sum_{u\in\Ucal} (1+\eps)P_U(u) I(W;V|U=u)+\alpha \log |\Xcal||\Ycal|\\
\vspace{5pt} \\
\label{rateclose} &=& (1+\eps)I(W;V|U)+\alpha \log |\Xcal||\Ycal|.
\end{IEEEeqnarray}
Step $(a)$ follows from $u^n(m)\in\Tcal^n_{\eps}$. 

The expression in \eqref{rateclose} is strictly less than $R_0$ for the proper choice of $\eps>0$ and $\alpha>0$. Thus, when $u^n(m)\in\Tcal_{\eps}^n$,
\begin{equation}
R_0 > \limsup_{n\to\infty} \frac1n i_{\Gamma}(W^n,X_{\Bcal},Y_{\Bcal};V^n).
\end{equation}
Invoking Lemma~\ref{generalcloudmixing}, we have
\begin{equation}
\lim_{n\to\infty}\Ebb _{\Ccal_{V}^{(n)}(m)} \Big\lVert Q_{W^nX_{\Bcal}Y_{\Bcal}|M=m} - \widehat{Q}_{W^nX_{\Bcal}Y_{\Bcal}|M=m} \Big\rVert =0.
\end{equation}
This completes the proof of Lemma~\ref{lemma:q_prop}.
\bibliographystyle{IEEEtran}
\bibliography{RDSS}
\end{document}